%% file: ms.tex
\documentclass[aps,prd,twocolumn,superscriptaddress,preprintnumbers,nofootinbib]{revtex4-2}

\usepackage{showyourwork}
\usepackage{amsfonts,amssymb,amsmath}
\usepackage[nolist,nohyperlinks]{acronym}
\usepackage{bookmark}

\newcommand{\infd}{\mathrm{d}}

\begin{document}

\title{Constraining gravitational wave amplitude birefringence with GWTC-3}

\author{Thomas C. K. Ng}
\email{thomas.ng@link.cuhk.edu.hk}
\affiliation{Department of Physics, The Chinese University of Hong Kong, Shatin, Hong Kong}

\author{Maximiliano Isi}
\email{misi@flatironinstitute.org}
\affiliation{Center for Computational Astrophysics, Flatiron Institute, 162 5th Avenue, New York, New York 10010, USA}

\author{Kaze W. K. Wong}
\email{kwong@flatironinstitute.org}
\affiliation{Center for Computational Astrophysics, Flatiron Institute, 162 5th Avenue, New York, New York 10010, USA}

\author{Will M. Farr}
\email{wfarr@flatironinstitute.org}
\affiliation{Center for Computational Astrophysics, Flatiron Institute, 162 5th Avenue, New York, New York 10010, USA}
\affiliation{Department of Physics and Astronomy, Stony Brook University, Stony Brook, New York 11794, USA}

\date{\today}

\begin{abstract}
    The propagation of gravitational waves can reveal fundamental features of the structure of spacetime.
    For instance, differences in the propagation of gravitational-wave polarizations would be a smoking gun for parity violations in the gravitational sector, as expected from birefringent theories like Chern-Simons gravity.
    Here we look for evidence of amplitude birefringence in the third catalog of detections by the Laser Interferometer Gravitational Wave Observatory and Virgo through the use of birefringent templates inspired by dynamical Chern-Simons gravity.
    From 71 binary-black-hole signals, we obtain the most precise constraints on gravitational-wave amplitude birefringence yet, measuring a birefringent attenuation of %
  \input{output/restricted_kappa_median.txt}\unskip\label{output/restricted_kappa_median.txt}\unskip%
 at $100 \, \mathrm{Hz}$ with 90\% credibility, equivalent to a parity-violation energy scale of %
  \input{output/M_PV_constraint.txt}\unskip\label{output/M_PV_constraint.txt}\unskip%
.
\end{abstract}

\maketitle

\begin{acronym}
\acro{GW}{gravitational wave}
\acro{GR}{general relativity}
\acro{CBC}{compact-binary coalescence}
\acro{BH}{black hole}
\acro{BBH}{binary black hole}
\acro{LVK}{LIGO-Virgo-KAGRA Collaboration}
\acro{PE}{parameter estimation}
\acro{FAR}{false-alarm rate}
\acro{SNR}{signal-to-noise ratio}
\end{acronym}

\section{Introduction}
\label{sec:Introduction}
\Ac{GW} detections by the \ac{LVK} \citep{LIGO, Virgo, KAGRA} are now routinely used to test various aspects of Einstein's theory of \ac{GR} \citep{LIGOScientific:2016lio,LIGOScientific:2018dkp,LIGOScientific:2021sio}.
Among those, measurements of the basic properties of \acp{GW}, like their speed and polarization, can directly probe the fundamental symmetries of the underlying theory of gravity \citep{Will:2018bme}.
For instance, unequal propagation of \ac{GW} polarization eigenstates would reveal that spacetime is birefringent, a smoking gun for parity-odd theories like Chern-Simons gravity \citep{Lue:1998mq,Jackiw:2003pm,Alexander:2009tp,Sopuerta:2009iy}.
Here we constrain the magnitude of possible amplitude birefringence using \ac{BBH} signals from the latest \ac{LVK} catalog, GWTC\nobreakdash-3 \citep{GWTC-3}.

Previous studies have constrained amplitude birefringence by performing different statistical analyses.
\citet{Yamada_2020}, \citet{Wang_2021} both performed \ac{PE} on the events in the first \ac{GW} transient catalog \citep{GWTC-1}, GWTC-1, using birefringent templates.
\citet{Okounkova_2022} considered the distribution of observed inclinations of the \ac{GW} events in the second \ac{GW} transient catalog \citep{GWTC-2}, GWTC-2, to look for signs of birefringence; \citet{Vitale:2022pmu} carried out a related analysis.
While this manuscript was finalized, \citet{Zhu:2023wci} reported an analysis of GWTC-3.

In this study, we use a frequency-dependent birefringence model to constrain the strength of \ac{GW} amplitude birefringence by performing \ac{PE} on \ac{LVK} binaries.
This model is a better approximation of the birefringence effect expected from theory rather than the frequency-independent model used in \citet{Okounkova_2022}.
Including the frequency dependence allows us to break the degeneracy between birefringence and source inclination, which we discuss below in Sec.~\ref{sec:inclination}.
Compared to Refs.~\citep{Yamada_2020,Wang_2021,Okounkova_2022}, we perform \ac{PE} on more events, including events new to GWTC-3 \citep{GWTC-3}, and use a phenomenology-oriented parametrization.
We consider 71 binary black hole merger events with a \ac{FAR} $\leq1/\mathrm{yr}$, as listed in Table I of \citet{GWTC-3_population}.
We discuss single-event results in detail, and identify degeneracies between birefringence and spin effects, in addition to the already known correlations with source orientation and distance.
We use the results from individual events to place a collective population constraint on the strength of \ac{GW} amplitude birefringence from GWTC-3.

In Sec.~\ref{sec:Background}, we briefly review the background of \ac{GW} amplitude birefringence.
In Sec.~\ref{sec:Method}, we describe the modification we apply to the baseline \ac{GR} waveform, summarize our \ac{PE} configuration and outline our statistical methods for combining information from multiple events.
In Sec.~\ref{sec:Results}, we present our constraint on \ac{GW} amplitude birefringence, discussing individual events and the catalog collectively.
In Sec.~\ref{sec:Discussion}, we discuss the implications of our result, comparing to previous constraints in the literature and outlining correlation structures that appear in our measurements.
Finally, we conclude in Sec.~\ref{sec:Discussion} with a summary.
In the Appendix, we provide extended results and discussion for two notable events (GW170818 and GW190521).

\section{Background}
\label{sec:Background}

\subsection{Birefringence}
\label{sec:waveform}

In \ac{GR}, \acp{GW} are comprised of two independent polarization modes, usually represented in the linear basis of plus ($+$) and cross ($\times$) states.
In the Fourier domain, these can be combined into left-handed (\textit{L}) and right-handed (\textit{R}) circular states (see, e.g., \cite{Isi:2022mbx}),
\begin{equation}
    h_{L/R} = \frac{1}{\sqrt{2}}\left(h_+ \pm i h_\times\right)\,,
\end{equation}
where $h$ is the frequency domain strain, with the plus and minus signs for L and R respectively.
These circular modes represent eigenstates of the helicity operator (helicity $\pm2$) and possess a definite parity.
Einstein's theory, which is parity even, predicts no difference in the dynamics of these two states.

Yet, parity odd extensions of \ac{GR} may make distinctions between the two circular polarizations, potentially appearing in both the generation and propagation of \acp{GW}.
The latter can manifest in changes to the relative amplitude and phase of the polarizations that accrue as the wave propagates, giving us hope of detecting initially small effects that compound over long propagation distances.

In particular, \emph{amplitude} birefringence would enhance one polarization mode over the other.
Following \cite{Alexander:2009tp}, the effect of birefringence can be modeled as a frequency-dependent amplification or dampening; similar derivations can be found in \cite{Zhao:2019xmm, Ezquiaga:2021ler, Zhu:2023wci,Jenks:2023pmk}.
To first order in the equations of motion, in theories like Chern-Simons gravity, the Fourier-domain waveform observed at a comoving distance $d_C$ away from the source can be written as
\begin{equation}
    h_{L/R}^{\mathrm{br}}(f) =
    h_{L/R}^{\mathrm{GR}}(f) \times
    \exp\left(\pm\kappa\, d_C \frac{f}{100\,\mathrm{Hz}}\right)\,,
    \label{eq:waveform_modification}
\end{equation}
where the emitted waveform $h_{L/R}^{\mathrm{GR}}$ is modified by an exponential birefringent factor to yield the observed waveform $h_{L/R}^{\mathrm{br}}$.
The overall magnitude of this effect for a given frequency $f$ is set by an attenuation coefficient $\kappa$, which encodes the intrinsic strength of the birefringence:
$\kappa^{-1}$ represents an ``attenuation length'' encoding the typical distance that yields an $e$-folding in the amplification or dampening of the polarizations at a fiducial signal frequency of 100 Hz at the detector.
The emitted waveform for a given source (i.e., the waveform observed in the near zone, very close to the source) will generally differ from the analogous waveform predicted by \ac{GR} \cite{Alexander:2009tp,Okounkova:2019zjf}; however, since we expect most viable modifications to \ac{GR} to be intrinsically small (e.g., \cite{Okounkova:2022grv}), it is standard to approximate the emitted waveform by the prediction from \ac{GR} (hence the notation ``$h^{\rm GR}$'' above).

Although the intrinsic modification is small, the effect targeted by Eq.~\eqref{eq:waveform_modification} accumulates as the \ac{GW} propagates.
During propagation, the effect of birefringence will be built up with the number of cycles, which is itself a function of the distance traveled and the frequency of the \acp{GW}.
According to Eq.~\eqref{eq:waveform_modification}, a positive $\kappa$ means the left-handed polarization is enhanced over the right-handed polarization, while a negative $\kappa$ means the opposite;
when $\kappa=0$, the observed waveform is the same as \ac{GR} predicts, meaning there is no birefringence.

Equation \eqref{eq:waveform_modification} can be derived as the first-order effect in an expansion away from \ac{GR} under multiple frameworks.
In general, $\kappa$ will be a function of the theory parameters and the cosmological history, e.g., the value of the pseudoscalar field and its derivative in Chern-Simons gravity \cite{Alexander:2009tp}.
Since it originates from a truncated series expansion,%
\footnote{Often the birefringent effect is written as an expansion in redshift $z$ \cite[e.g.][]{Zhao:2019xmm}, rather than distance. Here we use $d_C$ to emphasize that the effect accumulates \emph{per cycle}, or with propagation distance.  Thus, current ground-based experiments, which can detect gravitational waves to $z \simeq 1$, $d_C \simeq d_H$ (Hubble distance), are already observing propagation over a large fraction of the Universe and constrain this effect meaningfully.} %
Eq.~\eqref{eq:waveform_modification} is a good approximation only for small exponents, namely
\begin{equation}
    \left(\frac{\left|\kappa\right|}{\mathrm{Gpc}^{-1}}\right) \left(\frac{d_c}{\mathrm{Gpc}}\right) \left(\frac{f}{100\, \mathrm{Hz}}\right) < 1\,,
    \label{eq:small_exponent}
\end{equation}
recalling that $\kappa$ has dimensions of inverse length.
Otherwise, more frequency-dependent terms could enter the exponent of Eq.~\eqref{eq:waveform_modification} in a theory-dependent way.
\begin{figure}
    \script{birefringence.py}
    \includegraphics[width=\columnwidth]{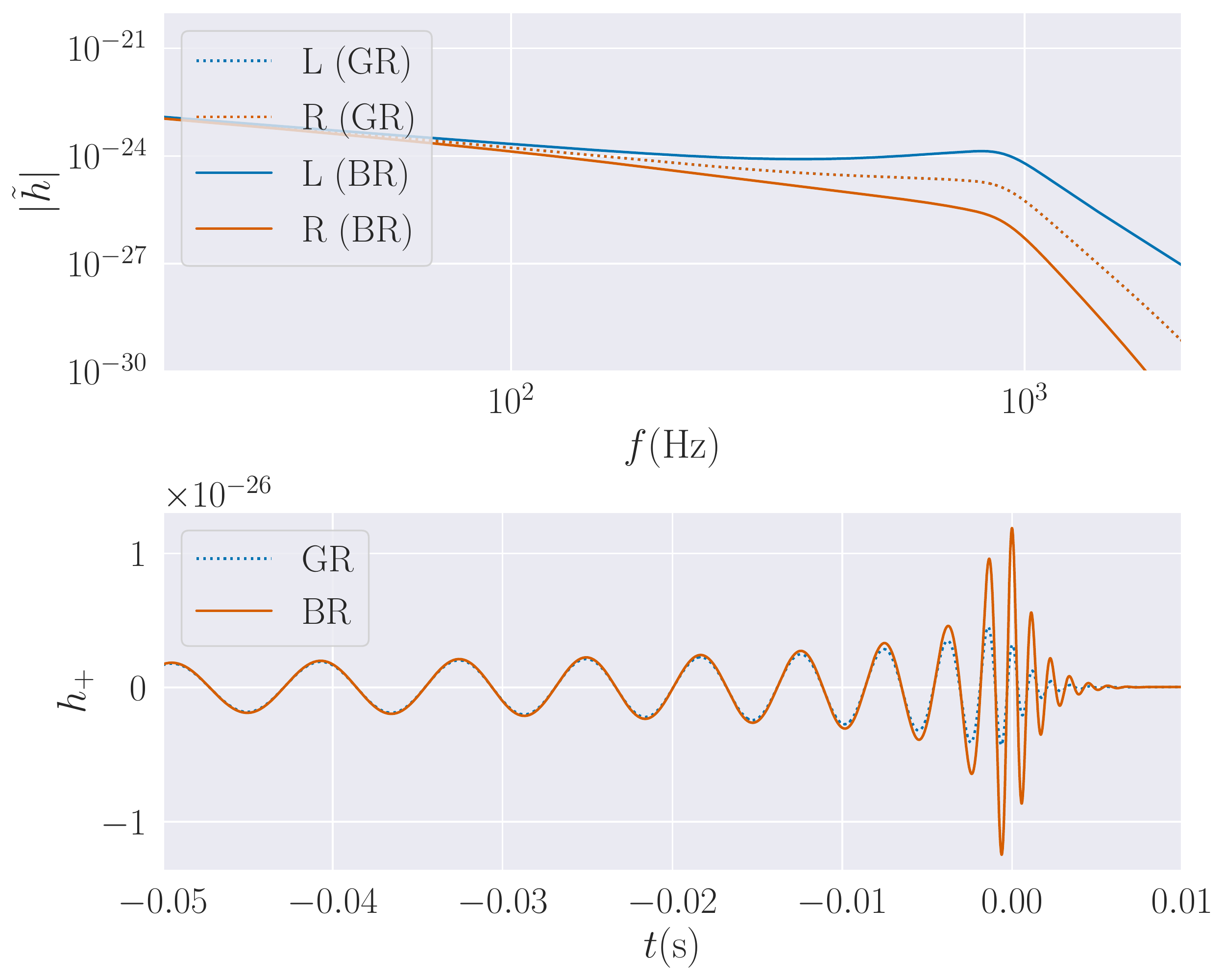}
    \caption{
        Illustration of amplitude birefringence. The GR waveform for the $\ell=|m|=2$ mode of a nonprecessing BBH seen edge-on ($\cos\iota = 0$) is linearly polarized and thus contains equal amounts of left- and right-handed modes for all frequencies (dotted, top).
        However, if spacetime were birefringent (BR) following Eq.~\protect\eqref{eq:waveform_modification}, then the waveform observed on Earth would contain different fractions of the two circular modes, with higher frequencies affected more strongly (solid, top). 
        In the time domain, this manifests as a time-dependent amplification of the waveform, with a stronger effect at later times when the chirp reaches a higher instantaneous frequency (bottom).
        For this example, the black holes do not spin and have equal masses $m_1 = m_2 = 10\, M_\odot$, and we have chosen a luminosity distance $d_L = 400\, {\rm Mpc}$ and $\kappa = 0.6$.
        }
    \label{fig:birefringence}
\end{figure}

\subsection{Inclination and other degeneracies}
\label{sec:inclination}

Under certain conditions, the effect of birefringence can be degenerate with a change in the orientation of the source with respect to the line of sight \cite{Alexander:2009tp}.
Concretely, for a nonprecessing compact binary inspiral in \ac{GR}, the observed amplitude ratio of the left-handed and right-handed polarizations is only a function of the inclination $\iota$, the angle between the orbital angular momentum of the source and the line of sight \cite{Blanchet:2013haa}.
For  the dominant $\ell = |m| = 2$ angular mode of the radiation, the relation between the amplitude ratio and the inclination is
\begin{equation}
    \frac{h_{L}^\mathrm{GR}}{h^\mathrm{GR}_{R}}=\left(\frac{1-\cos\iota}{1+\cos\iota}\right)^2\,
\end{equation}
for all frequencies (see, e.g., Sec.~III C in \cite{Isi:2022mbx}).

Since birefringence impacts the observed amplitude ratio of left- and right-handed modes, it could also affect inferences about the source inclination \cite{Alexander:2009tp}.
However, the two effects are degenerate only if the frequency dependence of Eq.~\eqref{eq:waveform_modification} is neglected.
This is easy to see from Eq.~\eqref{eq:waveform_modification} since the implied polarization ratio for the $\ell = |m| = 2$ mode of a nonprecessing source is
\begin{equation}
    \frac{h_{L}^\mathrm{br}}{h_{R}^\mathrm{br}}=\left(\frac{1-\cos\iota}{1+\cos\iota}\right)^2
    \exp\left(2\kappa d_C \frac{f}{100\, \mathrm{Hz}}\right)\, .
    \label{eq:modified_amplitude_ratio}
\end{equation}
For an isolated Fourier mode of definite frequency $f$, the effect of birefringence will be degenerate with a change in inclination; however, if multiple modes come into play, then no change in inclination alone can mask the effect of birefringence, which will affect the time domain waveform nontrivially (Fig.~\ref{fig:birefringence}).

\citet{Okounkova_2022} took the effect of birefringence to be independent of the frequency, which is a simplified approximation of the birefringence model in Chern-Simons gravity.
This assumption results in a full degeneracy between $\kappa$ and $\iota$:
to reconstruct the amplitude ratio from the interferometer data, a value of $\iota$ representing a more face-off inspiral can pair with a positive value of $\kappa$, or a value of $\iota$ representing a more face-on inspiral with a negative value of $\kappa$.
That fact can be used to constrain frequency-independent birefringence by searching for features in the distribution of inferred inclinations \cite{Okounkova_2022}.

By implementing Eq.~\eqref{eq:waveform_modification}, we generally break the degeneracy between birefringence and source orientation; this was also the case in the frequency-dependent relations studied in \cite{Yamada_2020,Wang_2021}.
Nevertheless, there exist systems for which the degeneracy cannot be broken, in practice, because not enough frequencies are available in the data.
This may also be the case for quasimonochromatic sources, like nonaxisymmetric pulsars or very light binaries, which are well approximated by a single Fourier mode.

As we will find in Sec.~\ref{sec:Results}, the effect of birefringence can be (partially) degenerate with other parameters besides source inclination.
Indeed, the frequency-dependent amplification or dampening of polarizations caused by $\kappa$ can sometimes be absorbed by changes in intrinsic parameters, like the spins, with concurrent adjustments to the source inclination and distance.
As a measure of the \ac{BH} spin magnitudes along the orbital angular momentum, we will use the effective spin parameter \cite{Damour:2001tu,Ajith:2009bn,Santamaria:2010yb}
\begin{equation}
\chi_{\rm eff} \equiv \frac{1}{1+q} \left(\chi_{1\parallel} + q\chi_{2\parallel}\right)
\end{equation}
where $\chi_{i\parallel}$ are the norms of the projections of the dimensionless spin vectors along the orbital angular momentum, and the mass ratio is $q \equiv m_2/m_1 \leq 1$.
We will study the interplay between birefringence and precession by focusing on the posterior of the precessing spin parameter $\chi_p$, defined as \cite{Schmidt:2014iyl}
\begin{equation}
\chi_p \equiv \max\left\{ \chi_{1\perp},\, k \chi_{2\perp}\right\} ,
\end{equation}
where $\chi_{i\perp}$ are the norms of the projections of the dimensionless spin vectors onto the orbital plane at a reference time, and $k \equiv q\left(4 q +3\right) / \left(4 + 3q\right)$.
We chart this and other approximate degeneracies as part of the results presented in Sec.~\ref{sec:Results}.


\section{Method}
\label{sec:Method}

\subsection{Single-event parameter estimation}

To constrain birefringence, we reanalyze events from GWTC-3 \citep{GWTC-2.1, GWTC-3} implementing Eq.~\eqref{eq:waveform_modification} to directly obtain a posterior on $\kappa$ from the strain of each event.
We analyze the 71 \acp{BBH} that were detected with $\mathrm{FAR} < 1/\mathrm{yr}$, less stringent than the typical \ac{LVK} threshold of $1/1000\,\mathrm{yr}$ \cite{LIGOScientific:2020tif,LIGOScientific:2021sio}.
The \ac{FAR} values are typically determined by \ac{GR} pipelines, which could down-rank signals beyond \ac{GR} \cite{LIGOScientific:2020tif,Chia:2020psj,tgrsel}; however, since detectability through matched-filtering is most sensitive to the phase, not the amplitude, we expect only a minor decrease in sensitivity to the kind of birefringent signals explored here.
To avoid extended computations on longer signals and considering these are generally at closer distances, we do not analyze systems involving neutron stars in this work.
We procure strain data from Gravitational Wave Open Science Center \citep{GWOSC}.

We estimate source parameters using a custom version of the \textsc{Bilby} software \citep{Bilby}, modified from the baseline version to apply Eq.~\eqref{eq:waveform_modification} for any \ac{GR} baseline waveform.
We take the \ac{PE} configuration in \citep{GWTC-2.1, GWTC-3, GWTC-2.1_dataset, GWTC-3_dataset} as a starting point, with \textsc{IMRPhenomXPHM} \citep{Pratten:2020ceb} as the reference waveform.
We apply a distance prior corresponding to a uniform distribution over comoving volume and source-frame time (see, e.g., Eq.~10 in \cite{LIGOScientific:2019zcs}), and set the prior on $\kappa$ to be uniform between $-1 \, \mathrm{Gpc}^{-1}$ and $1 \, \mathrm{Gpc}^{-1}$. 

For GW190521, we increase the maximum distance allowed by the prior to $1.5\times$ the original value in \cite{GWTC-2.1_dataset}, as the birefringence effect results in posterior support at larger distances.
For GW190720, we decrease the analysis segment from 16 s to 8 s, in order to accommodate missing data near the edges of the 16 s segment in Virgo.
Otherwise, the configuration is as in \citep{GWTC-2.1, GWTC-3, GWTC-2.1_dataset, GWTC-3_dataset}.

In order to validate our \ac{PE} implementation, we reproduce the \ac{LVK} \ac{PE} results obtained assuming \ac{GR} by enforcing $\kappa = 0$; this also has the advantage of producing \ac{GR} runs that are directly comparable to our birefringent runs.
All our \ac{PE} results, including the \ac{GR} validation runs, are published in \citet{dataset}.

\subsection{Collective analysis}

\subsubsection{Shared birefringence parameter}

In the most simplified scenarios, birefringence is a property of spacetime that is not intrinsic to any source or region in space. 
Consequently, we should expect $\kappa$ to take the same value for all signals, whether it vanishes or not.
Under this assumption, the posterior on $\kappa$ inferred collectively from all events is simply obtained from the product of the individual likelihoods, $p(d_i \mid\kappa) \propto p(\kappa \mid d_i)/p(\kappa)$, such that
\begin{equation}
    p(\kappa \mid \{d_i\})\propto p(\kappa) \prod_{i}\frac{p(\kappa \mid d_i)}{p(\kappa)}\,,
    \label{eq:restricted_posterior}
\end{equation}
where $d_i$ is the strain data for the $i$th event, and $p(\kappa)$ is the prior on $\kappa$; since the prior is uniform, in our case Eq.~\eqref{eq:restricted_posterior} reduces to the product of the posteriors, namely $p(\kappa \mid \{d_i\}) \propto \prod_{i}p(\kappa \mid d_i)$.
We use Eq.~\eqref{eq:restricted_posterior} to obtain the primary constraint presented in this work.

\subsubsection{Nonshared birefringence parameters}
\label{sec:method:hier}

Under many frameworks, birefringence is brokered by an extra parity-odd field that couples to gravity.
In that case, the effective strength of birefringence may vary along the different lines of sight to each event, depending on cosmic history and the local evolution of the field.
The \emph{simplest} assumption is that the field manifests equally for all events, as assumed in the previous subsection, but this is not strictly required.
These considerations motivate a collective analysis that does not assume $\kappa$ is shared across events \cite{Zimmerman:2019wzo,Isi:2022cii}.
Such an analysis has the additional advantage of helping us further characterize our set of measurements, and identify potential outliers.

To do this, we apply hierarchical Bayesian inference \cite{Mandel2010,Hogg2010} to model the distribution of $\kappa$'s consistent with the observed data:
we posit that, rather than a unique global value of $\kappa$, there is a specific value of the parameter, $\kappa_i$, associated with each event, and that this is drawn from some unknown distribution of true underlying values; from the imperfect measurements of $\kappa_i$ for each event, we may reconstruct the underlying distribution.
If we are interested in constraining the first two moments of the distribution, then it is convenient to parametrize the $\kappa_i$'s as drawn from a Gaussian with unknown mean $\mu$ and variance $\sigma^2$, i.e., $\kappa_i \sim \mathcal{N}(\mu, \sigma^2)$ \cite{Isi:2019asy}, and measure those hyperparameters from the collection of observed data.

If \ac{GR} is correct and there is no birefringence, then we should find the observed $\kappa$ distribution to be consistent with a delta function at the origin ($\kappa_i = 0$ for all $i$, or $\mu=\sigma=0$); on the other hand, if spacetime is globally birefringent, then we expect to find a delta function at some nonzero value ($\kappa_i = \kappa \neq 0$, or $\mu = \kappa$ and $\sigma=0$).
But this analysis also has the power to reveal unexpected physics or systematics in our measurements: if $\sigma$ is confidently found to be nonzero, then this would imply that our set of measurements is statistically unlikely to originate from a unique $\kappa$ value.
This could signal richer physics than is implied by Eq.~\eqref{eq:waveform_modification} or, more prosaically, that there are outliers in our measurements due to mismodeling, e.g., in the waveform approximant or the noise of the detector.

Starting from the posterior on $\kappa$ from each $i$\textsuperscript{th} event, $p(\kappa_i\mid d_i)$, the posterior on the hyperparameters $\mu$ and $\sigma$ can be calculated by
\begin{equation}
    p(\mu,\sigma \mid \{d\})\propto p(\mu,\sigma)\prod_{i}\int\frac{p(\kappa_i\mid d_i)}{p(\kappa_i)}p(\kappa_i\mid\mu,\sigma)\,\infd\kappa_i,
    \label{eq:posterior_of_mu_sigma}
\end{equation}
where $p(\kappa)$ is the prior initially applied to $\kappa$ during \ac{PE}, which in our case is a uniform distribution, $\mathcal{U}[-1,1]$.
Further choosing the hyperpriors on $\mu$ and $\sigma$ to be uniform, Eq.~\eqref{eq:posterior_of_mu_sigma} simplifies to
\begin{equation}
    p(\mu,\sigma\mid\{d\})\propto\prod_{i}\int p(\kappa_i\mid d_i)\, p(\kappa_i\mid\mu,\sigma)\,\infd\kappa_i\,
\end{equation}
where $p(\kappa_i\mid\mu,\sigma) \propto \exp(-|\kappa_i - \mu|^2/2\sigma^2)$ is the usual Gaussian likelihood.

Beyond measuring the population mean and variance, we also calculate the expected population distribution of $\kappa$ marginalized over $\mu$ and $\sigma$.
This is defined by
\begin{equation}
p(\kappa_i \mid \{d\})=\int p(\kappa_i \mid \mu,\sigma)p(\mu,\sigma\mid \{d\})\,\infd\mu\,\infd\sigma\, , 
    \label{eq:generic_posterior}
\end{equation}
and represents our overall expectation for the true values of $\kappa$, given the observed set of individual measurements and, crucially, assuming the underlying distribution is Gaussian.
Although the measurement of the population mean and variance applies irrespective of whether the underlying $\kappa_i$ distribution is truly Gaussian, the specific shape of the population-marginalized distribution of Eq.~\eqref{eq:generic_posterior} is not; therefore, Eq.~\eqref{eq:generic_posterior} should not be interpreted as giving a generic inference on the $\kappa$ distribution---more expressive models than a Gaussian would be better suited to that purpose. 
To sample the posterior distribution of $\mu$ and $\sigma$, we use the sampling package \textsc{flowMC} \citep{flowMC}.

\section{Results}
\label{sec:Results}

In this section, we present the results of our study.
We first show the $\kappa$ measurements from all events in our set, as well as the resulting global measurement of $\kappa$ that represents our primary constraint on birefringence (Sec.~\ref{sec:results:gwtc}).
We then assess the collection of measurements in more detail through a hierarchical analysis (Sec.~\ref{sec:results:hier}).
Finally, we discuss some special events, individually, and outline the degeneracies that arise between birefringence and orbital precession (Sec.~\ref{sec:results:notable}).

\begin{figure}
    \script{violin_kappa.py}
    \includegraphics[width=\columnwidth]{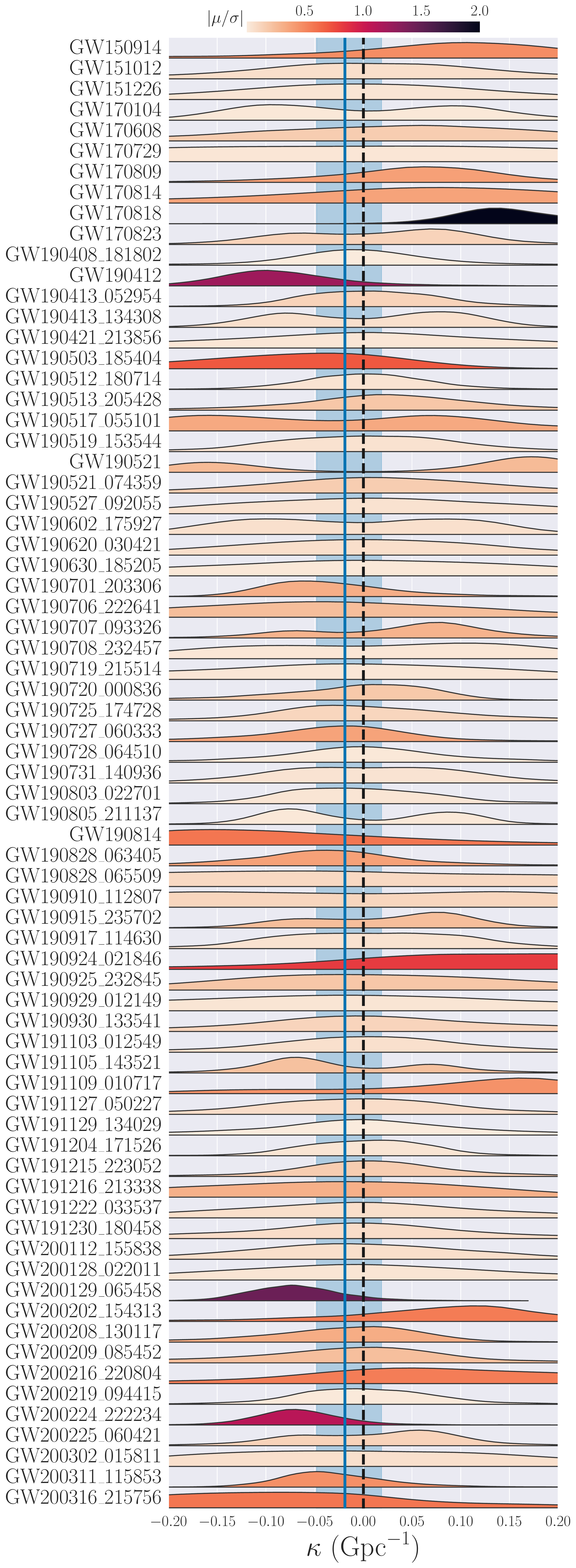}
    \caption{
        Individual-event $\kappa$ posteriors (distributions) and joint measurement (blue band, 90\% CI; blue line, median).
    }
    \label{fig:violin_kappa}
\end{figure}

\subsection{GWTC-3 result}
\label{sec:results:gwtc}

Figure \ref{fig:violin_kappa} shows the main result of our GWTC-3 analysis as represented by the posterior distribution on $\kappa$ (abscissa) obtained individually for each event (ordinate).
Posteriors are colored by the respective posterior mean distance to the origin in units of standard deviation, i.e., $|\mu_i/\sigma_i|$ for each event $i$.\footnote{We use the $i$ notation for individual events here to disambiguate from the population mean $\mu$ and standard deviation $\sigma$ of Eq.~\eqref{eq:posterior_of_mu_sigma}.}
The collective measurement obtained by assuming a shared $\kappa$ across events, Eq.~\eqref{eq:restricted_posterior}, is represented by its 90\%-credible symmetric interval (blue band) around the median (blue line); this joint measurement is fully consistent with $\kappa = 0$ (dashed line) with the credible level of %
  \input{output/CL_kappa_0.txt}\unskip\label{output/CL_kappa_0.txt}\unskip%
, meaning we find no evidence for birefringence.

Figure \ref{fig:violin_kappa} makes it clear that not all \acp{BBH} in GWTC-3 are equally informative about birefringence.
When considered individually, the events that best constrain $\kappa$ are listed in Table~\ref{tab:best_events_kappa}, in order of increasing standard deviation $\sigma_i$.
That table also shows the credible level (CL) at which the posterior supports $\kappa = 0$, whereby $\mathrm{CL} = 0$ ($\mathrm{CL} = 1$) means the posterior supports that value with high (low) probability.
This quantity represents the relative height of the probability density function at $\kappa = 0$, and is not tied to a symmetric interval, making it particularly useful in assessing bimodal posteriors.

Judging by $\mu_i/\sigma_i$, the two events that show the largest tension with $\kappa = 0$ are GW170818, for which %
  \input{output/GW170818_constraint.txt}\unskip\label{output/GW170818_constraint.txt}\unskip%
, and GW200129\_065458 (henceforth GW200129), for which %
  \input{output/GW200129_constraint.txt}\unskip\label{output/GW200129_constraint.txt}\unskip%
.
However, as we discuss in Sec.~\ref{sec:GW200129}, we have reason to think that the preference for $\kappa < 0$ in GW200129 might be driven by noise anomalies in the Virgo detector; with that in mind, in the next section we consider the effect of excluding this event from the joint result (we find its impact to be minimal).

\begin{table}
    \caption{Events that best constrain $\kappa$, sorted by posterior standard deviation $\sigma_i$. CL is the credible level of $\kappa = 0$.}
    \begin{ruledtabular}
  \input{output/best_events_kappa.txt}\unskip\label{output/best_events_kappa.txt}\unskip%

    \end{ruledtabular}
    \label{tab:best_events_kappa}
\end{table}

\begin{table}
    \caption{Events with bimodality in the $\kappa$ posterior, the \ac{GR} measurement of their detector-frame total mass ($M$), precessing spin $\chi_p$, and effective spin $\chi_{\rm eff}$, as well as the CL of $\kappa = 0$ from the birefringence analysis.}
    \begin{ruledtabular}
  \input{output/bimodal_events_mass.txt}\unskip\label{output/bimodal_events_mass.txt}\unskip%

    \end{ruledtabular}
    \label{tab:bimodal_events_mass}
\end{table}

Finally, a set of events stands out in Fig.~\ref{fig:violin_kappa} due to evident bimodality in the $\kappa$ posterior.
To a varying degree, that is the case for those events listed in Table~\ref{tab:bimodal_events_mass}, which tend to have quite high total masses in the detector frame (Table~\ref{tab:bimodal_events_mass} shows total mass as measured in the standard \ac{GR} analysis).
For these bimodal posteriors, $\mu_i/\sigma_i$ is not a good proxy for agreement with \ac{GR}; instead, we can rely on $\mathrm{CL}(\kappa = 0)$.
By this measure, the bimodal events are some of the least consistent with $\kappa = 0$, GW190521, in particular.

As we anticipated in Sec.~\ref{sec:inclination}, we understand the bimodality in $\kappa$ to be linked to spin effects and often to precessing morphologies in particular.
Other parameter degeneracies also come into play, especially for the lighter events GW191105\_143521 (henceforth GW191105) and GW170104.
We discuss this further in a dedicated section below (Sec.~\ref{sec:results:notable}).

\begin{figure}
    \script{corner_Gaussian.py}
    \includegraphics[width=\columnwidth]{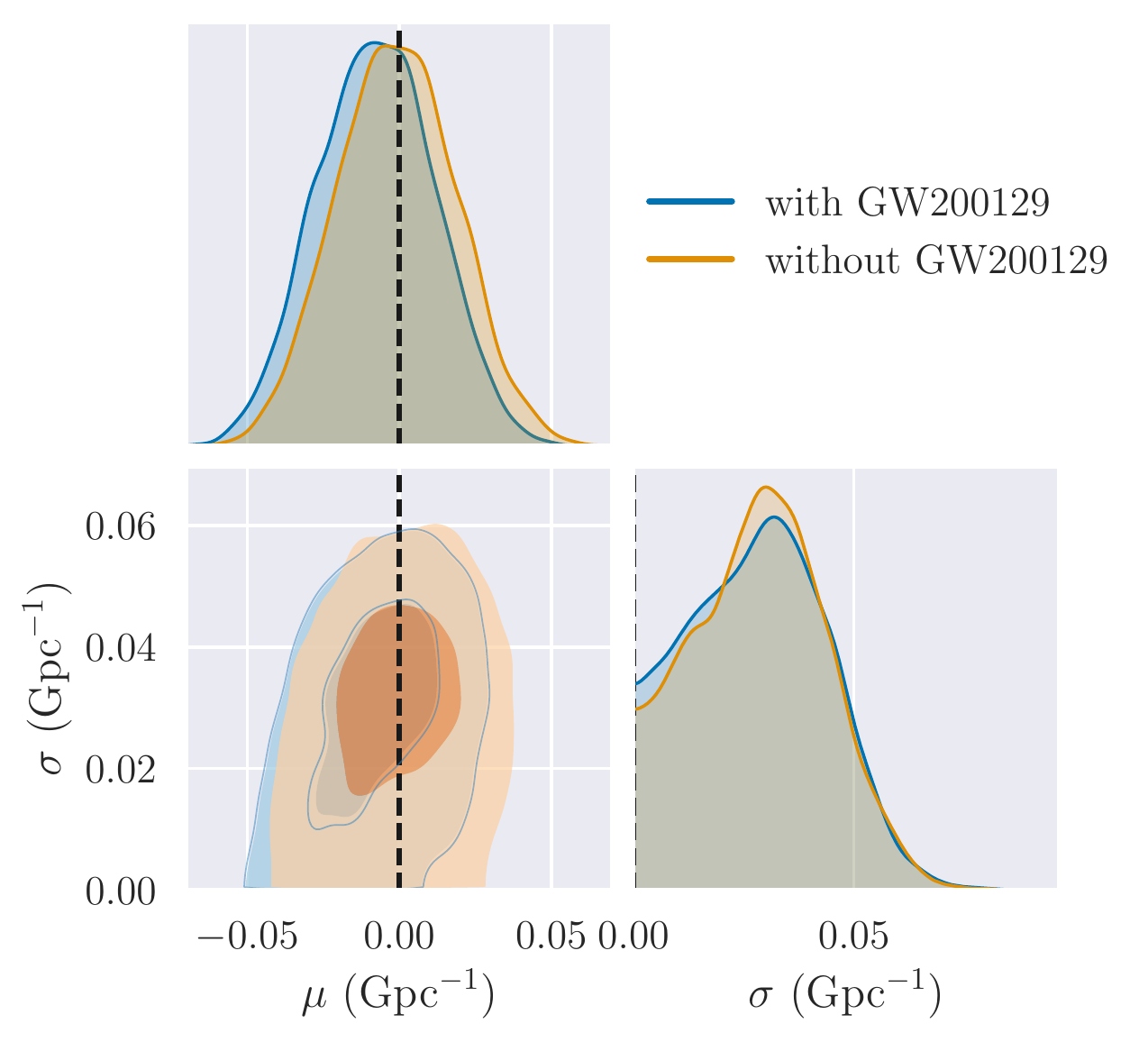}
    \caption{
        The posterior of the $\kappa$ population hyperparameters $\mu$ and $\sigma$, including (blue) and excluding (orange) GW200129 from the collection of events.
        The two-dimensional (2D) contours correspond to the $39.35\%$ and $90\%$ credible levels.
        The plot shows that the population constraint on $\kappa$ is consistent with no birefringence ($\mu=\sigma=0$) at the 90\% credible level.
    }
    \label{fig:corner_Gaussian}
\end{figure}

\subsection{Hierarchical modeling}
\label{sec:results:hier}

Figure \ref{fig:violin_kappa} shows a certain degree of variance in the distribution of $\kappa$ posteriors for different events, including some apparent outliers like GW170818 or GW190521.
This is not unexpected: assuming independent Gaussian noise instantiations for each event, we might expect up to ${\sim}3$ out of the 71 posteriors (i.e., ${\sim}5\%$) to deviate away from the true $\kappa$ value by over ${\sim}2\sigma$ due to noise alone.

To further assess the statistical properties of the set of posteriors in Fig.~\ref{fig:violin_kappa}, we apply the hierarchical analysis described in Sec.~\ref{sec:method:hier}.
By characterizing the population mean and standard deviation over events, this also allows us to obtain a collective measurement that does not assume all events share the same value of $\kappa$.

\begin{figure}
    \script{reweighed_kappa.py}
    \includegraphics[width=\columnwidth]{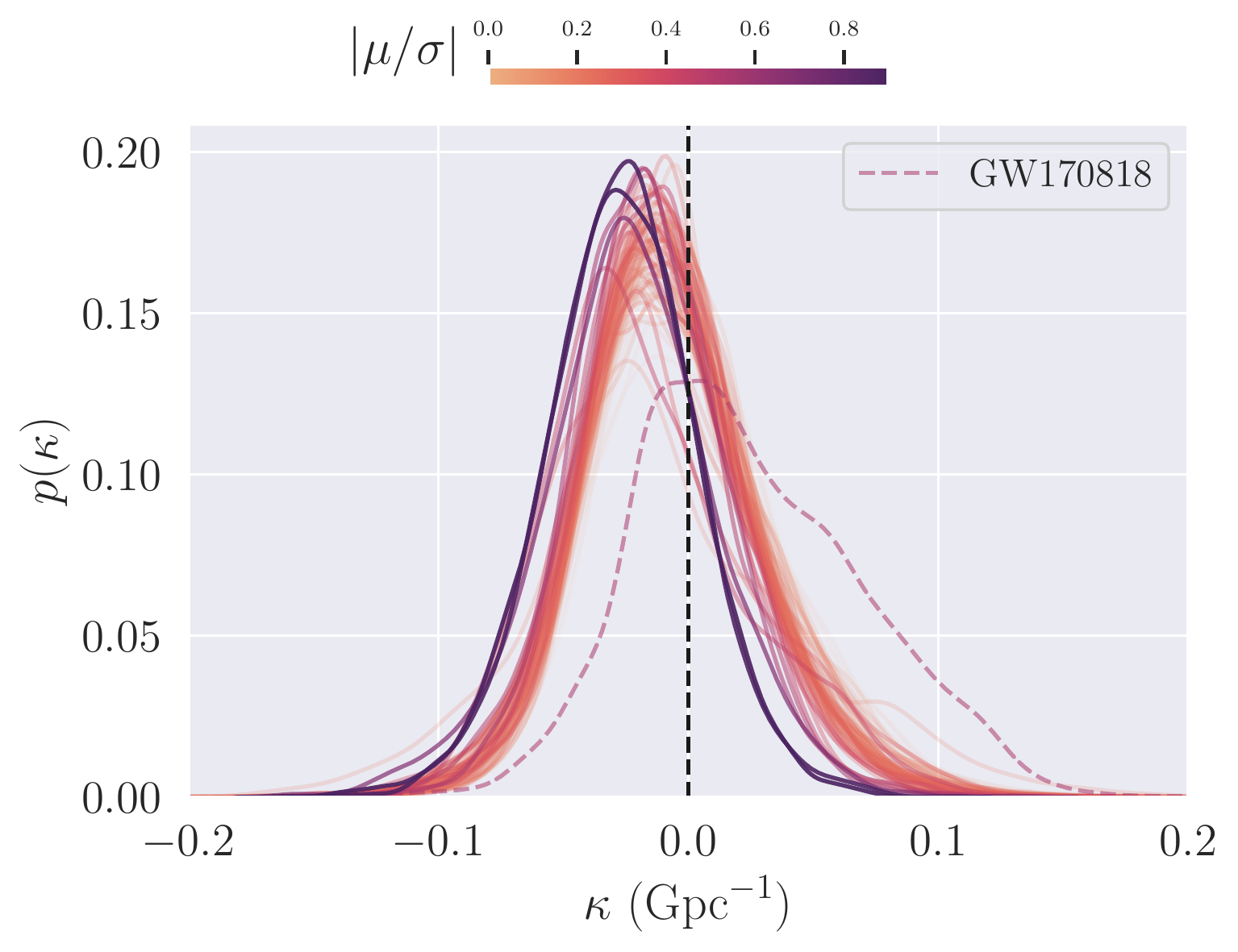}
    \caption{
        Individual-event $\kappa_i$ distributions of Fig.~\ref{fig:violin_kappa} reweighted in light of the population-level inference of Fig.~\ref{fig:corner_Gaussian}, marginalizing over $\mu$ and $\sigma$.
        Curves are colored by the magnitude of the supported deviation, as represented by the respective posterior $|\mu_i / \sigma_i|$.
    }
    \label{fig:reweighted_kappa}
\end{figure}

We summarize the result of this exercise in Fig.~\ref{fig:corner_Gaussian}, which shows the posterior on the population mean $\mu$ and standard deviation $\sigma$ inferred from the collection of observations in Fig.~\ref{fig:violin_kappa}.
The figure shows two distributions, which result from analyses with (blue) and without (orange) the potentially-contaminated event GW200129; the difference between the two is mainly restricted to a slight shift in $\mu$, indicating that GW200129 has a small effect on our overall population inference.

Both versions of the posterior support the lack of birefringence ($\mu = \sigma = 0$) within 90\% credibility; from the marginals of the result including all events, we constrain %
  \input{output/mu_median.txt}\unskip\label{output/mu_median.txt}\unskip%
for 90\%-credible symmetric intervals around the median, and %
  \input{output/sigma_median.txt}\unskip\label{output/sigma_median.txt}\unskip%
for the 90\%-credible one-sided upper limit.
However, even though $\sigma = 0$ is well supported, the $\sigma$ posterior peaks visibly away from the origin, indicating some preference for a nonzero variance.
This could be a sign of the presence of outliers in our sample.

As a visual check for outliers, we reconsider the set of measurements in Fig.~\ref{fig:violin_kappa} in light of the hierarchical result for $\mu$ and $\sigma$ in Fig.~\ref{fig:corner_Gaussian}, including GW200129 (blue curve).
This amounts to reweighting the $\kappa$ posterior for each event under a population prior marginalized over $\mu$ and $\sigma$, conditional on the measurements from all other events \cite{Miller2020,Callister:T2100301}.
The result in Fig.~\ref{fig:reweighted_kappa} does not show evidence for any of the events being in obvious tension with the population, even though the GW170818 curve stands out from the rest due to its higher support for $\kappa > 0$.
This feature appears to offset a few other events which tend to favor $\kappa < 0$.
The interaction between these distributions leads to a hyperposterior that is fully consistent with $\mu = 0$ while offering some support for $\sigma > 0$ (Fig.~\ref{fig:corner_Gaussian}).
Future observations will determine whether there is truly evidence for a nonvanishing variance in this population.

\begin{figure}
    \script{posterior_kappa.py}
    \includegraphics[width=\columnwidth]{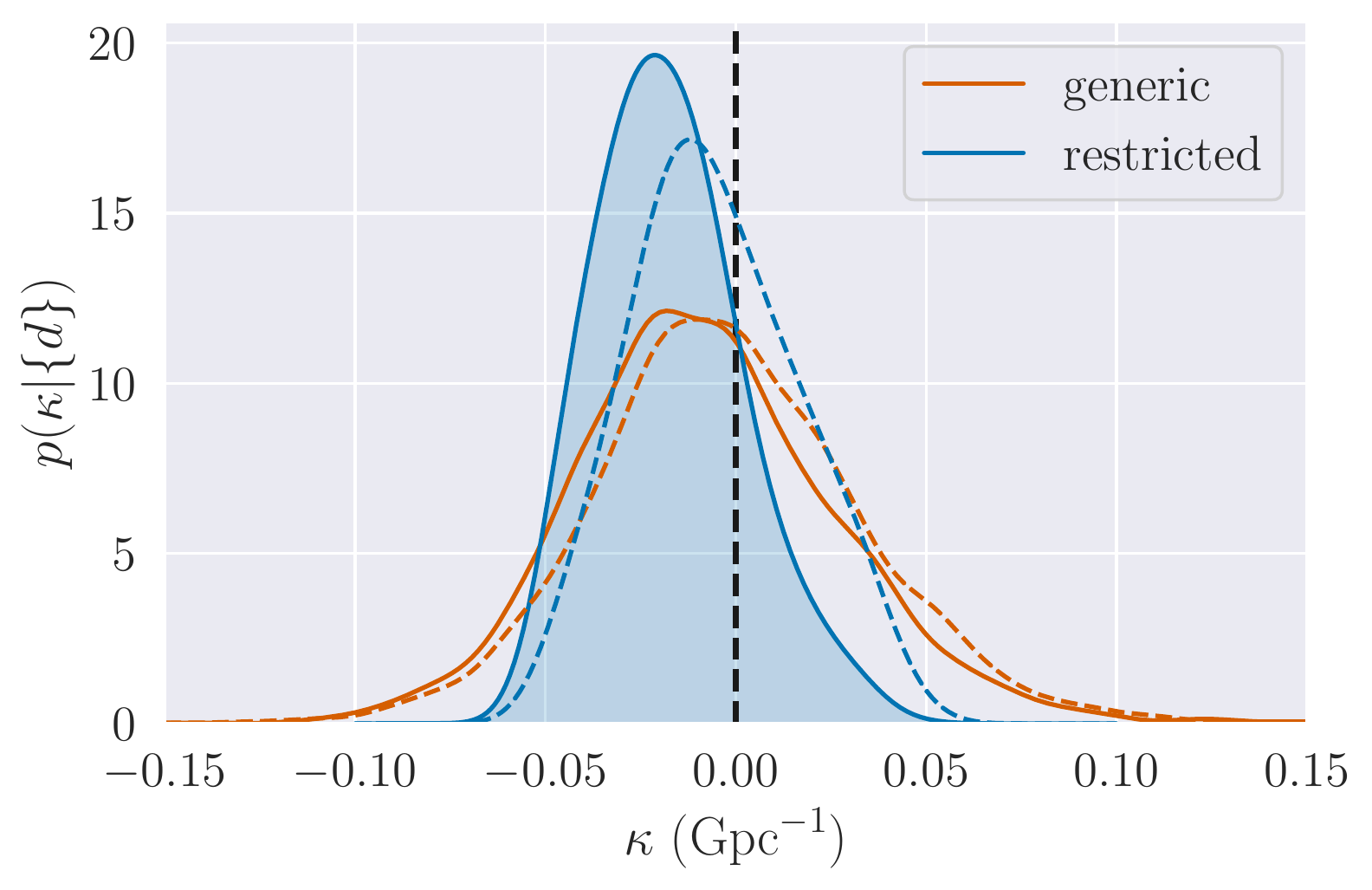}
    \caption{
        Restricted and generic posteriors on $\kappa$, which, respectively, do and do not assume that $\kappa$ is shared by all events (color).
        Solid (dashed) traces indicate the analysis included (excluded) GW200129.
        The black dashed line marks the absence of birefringence ($\kappa=0$).
        The shaded distribution is the primary result in this work (blue band in Fig.~\ref{fig:violin_kappa}).
    }
    \label{fig:posterior_kappa}
\end{figure}

Finally, the hierarchical result in Fig.~\ref{fig:corner_Gaussian} can be translated into an overall expectation for $\kappa_i$ under the assumption of a Gaussian distribution via Eq.~\eqref{eq:generic_posterior}.
We show the result of doing this in Fig.~\ref{fig:posterior_kappa}, where we also compare to the posterior on $\kappa$ obtained by assuming a shared value across events (same result shown as a blue band in Fig.~\ref{fig:violin_kappa}).
The hierarchical measurement leads to an expectation that %
  \input{output/generic_kappa_median.txt}\unskip\label{output/generic_kappa_median.txt}\unskip%
, whereas the shared-$\kappa$ measurement implies %
  \input{output/restricted_kappa_median.txt}\unskip\label{output/restricted_kappa_median.txt}\unskip%
, both symmetric 90\%-credible intervals around the median.  

\subsection{Notable events}
\label{sec:results:notable}

Having established that the collection of detections is globally consistent with $\kappa=0$, here we focus on three events whose $\kappa$ posteriors stand out in Fig.~\ref{fig:violin_kappa}: GW170818, GW190521, and GW200129.
When considered in isolation, the first of these is the unimodal event with the most significant support for nonzero $\kappa$, the second is the most extreme representative of a class of events with bimodal $\kappa$ posteriors, and the third shows signs of potential noise anomalies.
Through these examples, we elucidate some of the interactions between $\kappa$ and the source luminosity distance, inclination, and spin parameters.

\subsubsection{GW170818}
\label{sec:GW170818}

\begin{figure}
    \script{corner_GW170818.py}
    \includegraphics[width=\columnwidth]{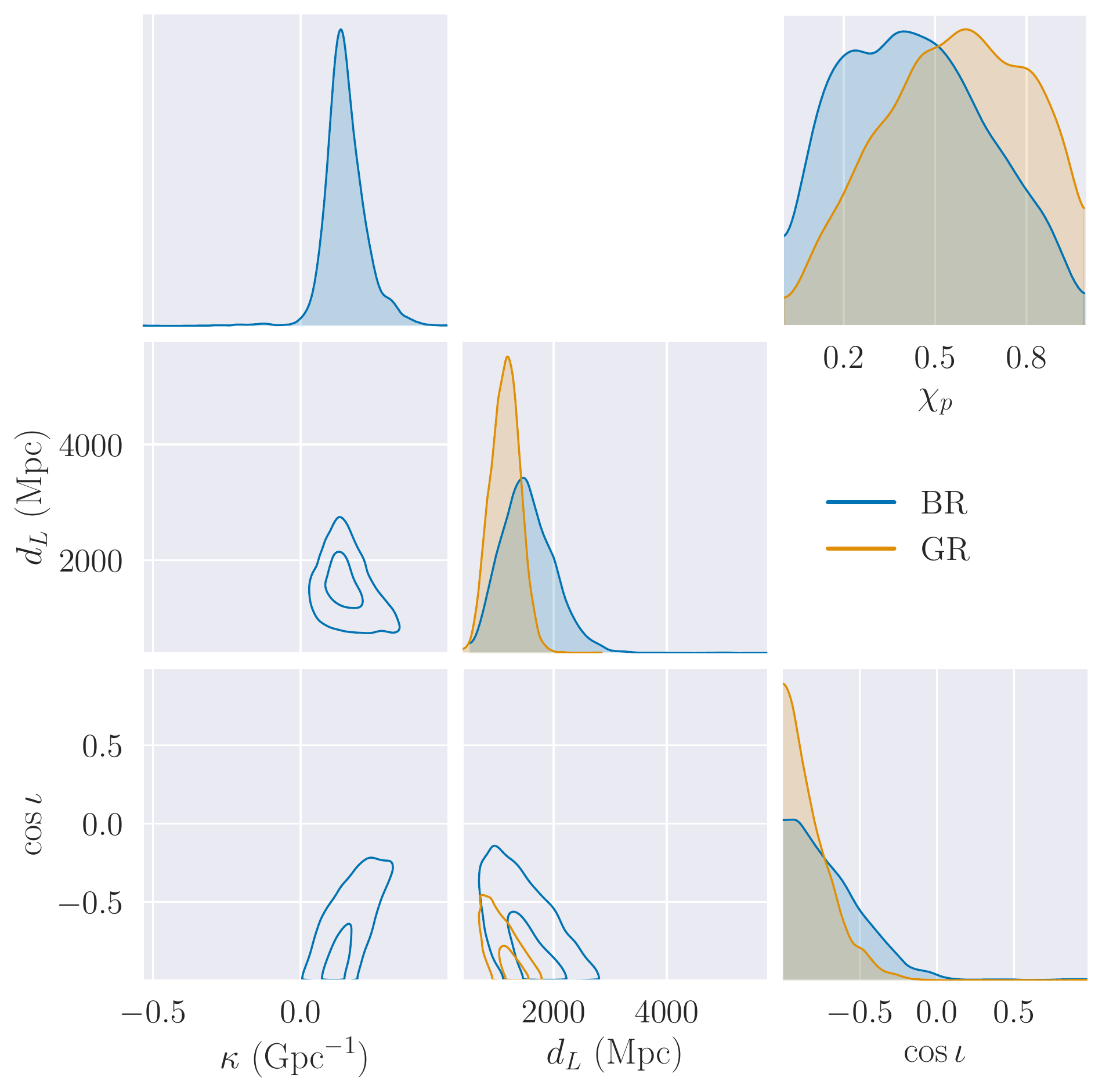}
    \caption{
        GW170818 posterior on $\kappa$, luminosity distance $d_L$, and inclination $\cos\iota$ from our birefringence analysis (blue), compared to the GR result (orange).
        The top right panel shows the marginalized posterior on $\chi_p$: allowing for birefringence reduces the preference for precession. (See Fig.~\ref{fig:corner_GW170818_appendix} for a full corner plot.)
    }
    \label{fig:corner_GW170818}
\end{figure}

GW170818 produced the posterior most displaced from $\kappa=0$, when judged by $\mu_i/\sigma_i$ in Fig.~\ref{fig:violin_kappa}.
Figure \ref{fig:corner_GW170818} shows that this happens because birefringence opens up a region of parameter space with $\kappa>0$ for larger distances and smaller inclination angles than would be allowed in the GR case.
We can make sense of this by noting that an edge-on nonprecessing source produces linearly-polarized waves, meaning that a smaller inclination leads the two circular polarizations to have similar amplitudes.
On the other hand, having $\kappa > 0$ enhances the left-handed modes during propagation, per Eq.~\eqref{eq:waveform_modification}.
The two effects can be balanced to match the polarization ratio observed at the detector (predominantly left handed, per the preference for $\cos\iota \approx -1$ in the \ac{GR} analysis), as long as the distance is also enhanced to yield the right amount of birefringence and overall signal power.
This is similar to the degeneracy mentioned in Sec.~\ref{sec:inclination}.

It is difficult to unequivocally identify a specific feature of the GW170818 data that leads to this posterior structure.
However, it appears to be related to this event's support for precession, in conjunction with its uncommonly definite measurement of the polarization, phase, and spin angles \cite{Varma:2021csh} (see Appendix \ref{sec:corner_GW170818_appendix}).
The relevance of precession is evident from the posterior on $\chi_p$ (Fig.~\ref{fig:corner_GW170818}, top right): allowing for $\kappa \neq 0$ leads to reduced support for precession.
We can understand this in reference to Fig.~\ref{fig:birefringence}: if only a short portion of the signal is observed, then the frequency-dependent signal enhancement or dampening due to birefringence can mimic the time-dependent amplitude modulation produced by a precession cycle.
Therefore, under these circumstances, similar morphologies can be obtained by setting $\chi_p > 0$ or $\kappa > 0$, as long as the distance and inclination can also be adjusted accordingly.

The fact that the birefringent analysis favors a high $\kappa$ rather than a high $\chi_p$ can be explained by a consequence of prior volume: many more configurations are available with long distances and high $\kappa$ than with short distances and small $|\kappa|$.
The preference for $\kappa > 0$ over $\kappa < 0$ (and, therefore, the lack of bimodality in the $\cos\iota$ and $\kappa$ posteriors), is likely related to both the definite measurement of left-handed polarizations and the specific phasing of the precession cycle.
The latter manifests as a precise constraint on the spin orientation and phase angles in the \ac{GR} analysis \cite{Varma:2021csh} (see also Appendix \ref{sec:corner_GW170818_appendix}).
The observed amplitude modulation (say, increasing vs decreasing towards the merger) likely determines the allowed sign of $\kappa$ for this event.
We provide the full corner plot for all relevant parameters in Fig.~\ref{fig:corner_GW170818_appendix} in Appendix~\ref{sec:corner_GW170818_appendix}.

\subsubsection{GW190521}
\label{sec:GW190521}

\begin{figure}
    \script{corner_GW190521.py}
    \includegraphics[width=\columnwidth]{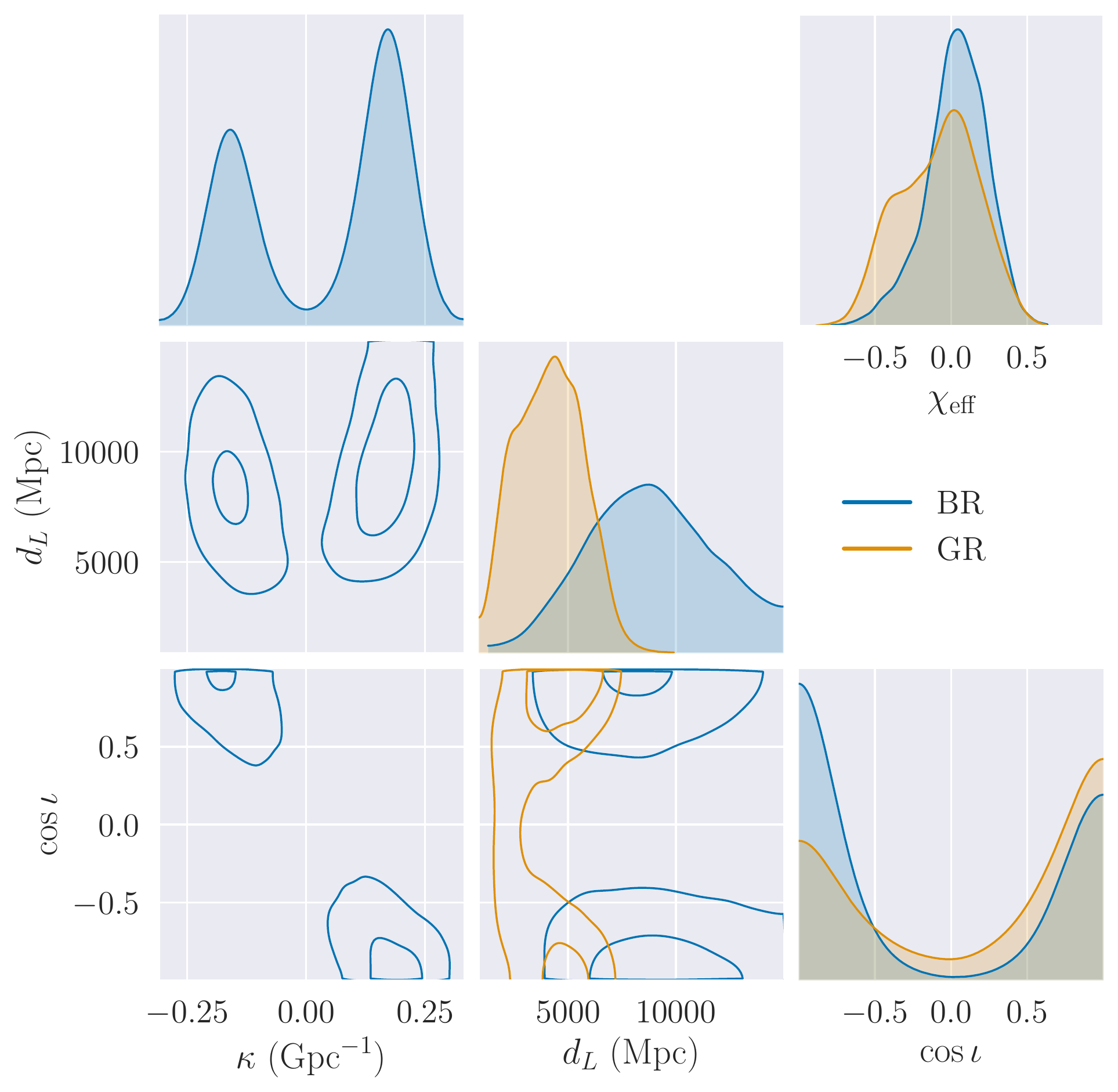}
    \caption{
        GW190521 posterior on $\kappa$, luminosity distance $d_L$, and inclination $\cos\iota$ from our birefringence analysis (blue), compared to the GR result (orange).
        The marginalized posterior on $\chi_p$ is shown in the top right panel. (See Fig.~\ref{fig:corner_GW190521_appendix} for a full corner plot.)
        Note that, although we show $d_L$ in this plot, our definition of $\kappa$ is associated with $d_C$, meaning that the condition in Eq.~\eqref{eq:small_exponent} is satisfied even for the largest values of $d_L$ supported in this posterior.
    }
    \label{fig:corner_GW190521}
\end{figure}

GW190521 is the most extreme representative of a class of events with bimodal $\kappa$ posteriors (Table \ref{tab:bimodal_events_mass}).
Figure \ref{fig:corner_GW190521} shows that the two peaks of the $\kappa$ distribution arise from respective modes in the inclination (lower left corner): one solution corresponds to an intrinsically face-on source ($\cos\iota \approx 1$) with negative $\kappa$, and the other to a face-off source ($\cos\iota \approx -1$) with positive $\kappa$.
In the former, an initially right-handed signal is birefringently enhanced by $\kappa < 0$ as it propagates towards the detector; in the latter, an initially left-handed signal is enhanced by $\kappa > 0$.
The two outcomes (a right-handed or left-handed signal at the detector) match the two possibilities allowed also in the reference \ac{GR} run, which itself yields a bimodal $\cos\iota$ posterior.
In both cases, the overall amplification of the signal introduced by a nonzero $\kappa$ is balanced by a large luminosity distance ($d_L \gtrsim 5 \, {\rm Gpc}$), so as to match the observed signal-to-noise ratio.

Introducing birefringence affects our inference of the spins in the GW190521 system.
We illustrate this in Fig.~\ref{fig:corner_GW190521} through the posterior on the effective spin $\chi_{\rm eff}$, which is favored to take values closer to zero in the birefringent analysis;
this can also be seen, perhaps more clearly, in the joint posterior for the individual spin magnitudes, provided in Appendix \ref{sec:corner_GW190521_appendix}.
This suggests that at least part of the data features that lead the \ac{GR} analysis to infer near extremal spins can be alternatively explained as a consequence of birefringence.
The fact that this is possible is not fully surprising given that the signal from this heavy system is only observed for a very small number of cycles \cite{LIGOScientific:2020iuh,LIGOScientific:2020ufj} (cf.~Fig.~\ref{fig:birefringence}).

All in all, the situation for GW190521 is not unlike that for GW170818 discussed above (Sec.~\ref{sec:GW170818}): structures in the data that would normally be interpreted as traces of high component spins are instead (at least partially) attributed to birefringent propagation, driven by a prior preference for high distances.
The degeneracies in the case of GW190521 are likely exacerbated by the heavier mass of the system, which reduces the observed number of cycles.
Crucially, GW190521 also lacks a definite measurement of the polarization state at the detector: without certainty about the ratio of right- versus left-handed components in the observed signal, it is not possible to disambiguate between $\kappa > 0$ versus $\kappa < 0$; hence the $\kappa$ distribution remains bimodal for GW190521, unlike for GW170818.
The poor determination of the polarization state of the detected signal is evidenced by the bimodal $\cos\iota$ posterior inferred assuming \ac{GR} in Fig.~\ref{fig:corner_GW190521}, and is attributable to several factors including sky location.

\subsubsection{GW200129}
\label{sec:GW200129}

GW200129 is the event with the second largest $|\mu_i/\sigma_i|$ in Fig.~\ref{fig:violin_kappa}.
However, data for this event were affected by a non-Gaussian noise disturbance (glitch) in the Virgo instrument, which was subtracted from the publicly-available data used for parameter estimation \cite{Davis:2022ird}.
Since previous work suggests the degree of glitch subtraction affects the inference for this event \citep{GW200129_glitch}, we consider whether the apparent preference for $\kappa < 0$ could also be tied to the instrumental artifact.

To this end, we perform three additional \ac{PE} runs for GW200129, considering only two detectors at a time: LIGO Hanford and Virgo (HV), LIGO Livingston and Virgo (LV), and LIGO Hanford and LIGO Livingston (HL).
If the preference for $\kappa < 0$ is tied to the glitch in Virgo, we expect it to disappear in the HL run, which excludes Virgo data.

\begin{figure}
    \script{corner_GW200129.py}
    \includegraphics[width=\columnwidth]{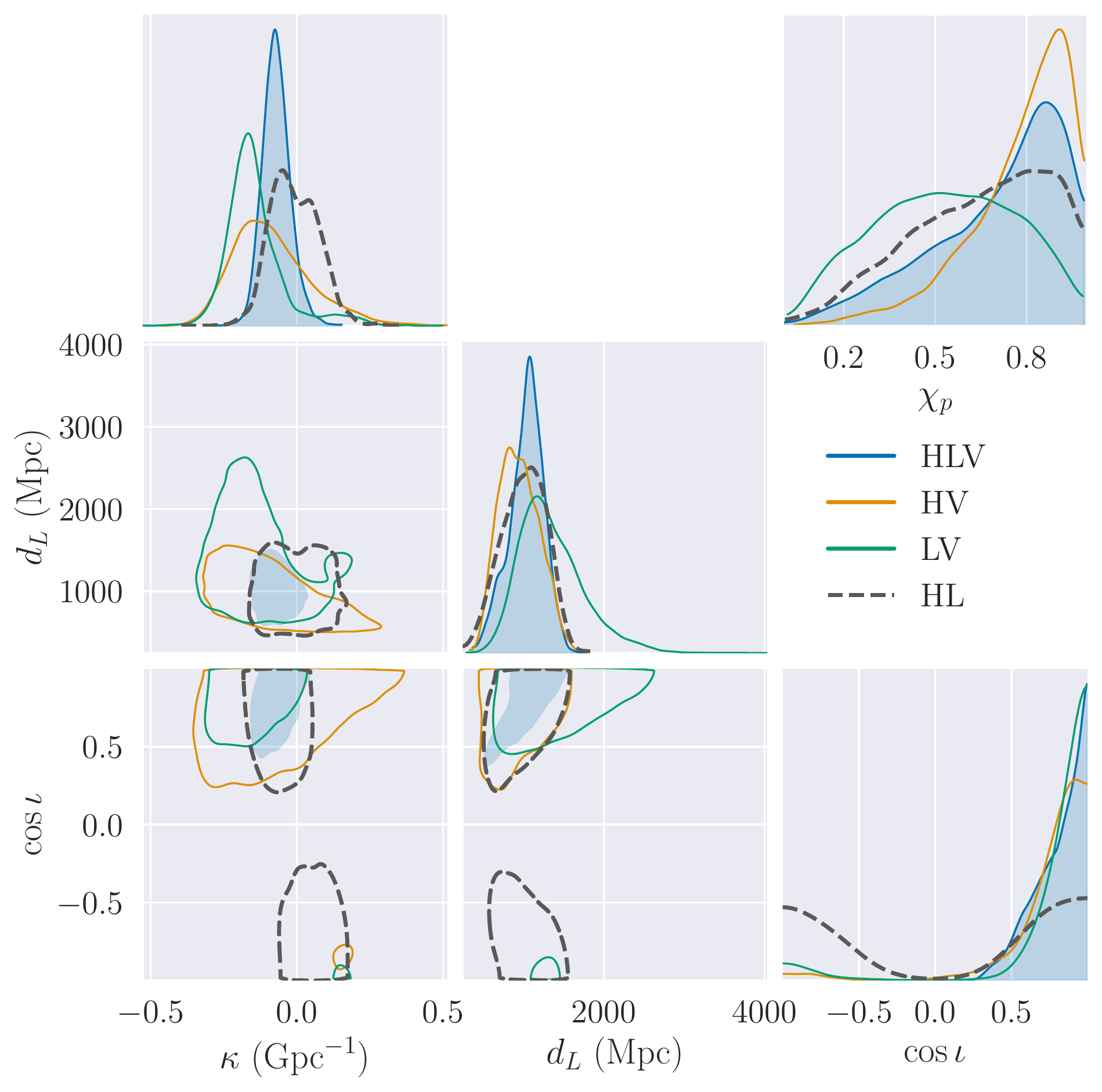}
    \caption{
        GW200129 posterior on $\kappa$, luminosity distance $d_L$, and inclination $\cos{\iota}$, including different sets of detectors in the analysis per the legend.
        The marginalized posterior on $\chi_p$ is shown in the top right panel.
        The main run with all three detectors (HLV, filled blue), shows a preference for $\kappa < 0$, as in Fig.~\ref{fig:violin_kappa}; this preference is more pronounced for two-detector runs that include Virgo (HV and LV, orange and green); however, it disappears if we remove Virgo (HL, dashed black).
        The 2D contours correspond to the $90\%$ credible level.
    }
    \label{fig:corner_GW200129}
\end{figure}

This is indeed the case, as we show in Fig.~\ref{fig:corner_GW200129}: all runs including Virgo lean towards $\kappa < 0$ (solid curves in color), whereas the LIGO-only run is fully consistent with $\kappa = 0$ (dashed black).
While this is not conclusive proof that the glitch itself is driving the result, it does indicate that the Virgo data play a key role in the inference of $\kappa$.
Since more work would be needed to understand the effect of the glitch, this motivated us to consider the effect of excluding this event from the collective analyses above (Sec.~\ref{sec:results:hier}).
In any case, the outcome of that test showed the role of GW200129 to not be significant (e.g., Fig.~\ref{fig:corner_Gaussian}).

\section{Discussion}
\label{sec:Discussion}

\subsection{Comparison with previous studies}

Our primary constraint on birefringence is given by the combined constraint on the birefringent attenuation from GWTC-3, %
  \input{output/restricted_kappa_median.txt}\unskip\label{output/restricted_kappa_median.txt}\unskip%
with 90\% credibility, obtained in Sec.~\ref{sec:results:gwtc} (blue band in Fig.~\ref{fig:violin_kappa}).
We can compare this to previous measurements of amplitude birefringence in the literature.
We summarize these constraints in Table~\ref{tab:comparison_summary}.

\begin{table}
    \caption{
        Comparison of our constraints with previous studies.
        CL is the credible level.
    }
    \begin{ruledtabular}
  \input{output/comparison_summary.txt}\unskip\label{output/comparison_summary.txt}\unskip%

    \end{ruledtabular}
    \label{tab:comparison_summary}
\end{table}

\subsubsection{Okounkova et al.}
\citet{Okounkova_2022} produced a constraint on frequency-independent \ac{GW} amplitude birefringence from the distribution of measured inclinations of \acp{BBH} in GWTC-2.
As opposed to our Eq.~\eqref{eq:waveform_modification}, this reference parametrized the effect of birefringence as
\begin{equation}
\label{eq:freqindep}
    h_{L/R}^{\mathrm{br}}(f) =
    h_{L/R}^{\mathrm{GR}}(f) \times
    \exp\left(\pm \tilde{\kappa}\, d_C\right)\,,
\end{equation}
where $\tilde{\kappa}$ is a frequency-independent attenuation (simply denoted ``$\kappa$'' in Ref.~\cite{Okounkova_2022}).
Assuming a fiducial redshift of $z=0.3$ for all events in GWTC-2, \citet{Okounkova_2022} obtained a constraint of $|\tilde{\kappa}| \lesssim 0.74\, {\rm Gpc}^{-1}$ at 68\% credibility.
On the other hand, our constraint from Sec.~\ref{sec:results:gwtc} is %
  \input{output/restricted_absolute_kappa_68.txt}\unskip\label{output/restricted_absolute_kappa_68.txt}\unskip%
 at $100 \, \mathrm{Hz}$ at 68\% credibility.
This is a factor of %
  \input{output/improvement_Okounkova.txt}\unskip\label{output/improvement_Okounkova.txt}\unskip%
more stringent than Ref.~\cite{Okounkova_2022}, besides arising from a less simplified model.

The result of \citet{Okounkova_2022} is also phrased in terms of a canonical Chern-Simons length scale,\footnote{This results from setting the so-called ``canonical'' Chern-Simons embedding into a Friedmann-Robertson-Walker cosmology, where the Chern-Simons field evolves as $\theta \propto t$ \cite{Alexander:2009tp,Jackiw:2003pm,Yunes2009}.}
\begin{equation}
    l_0 = \frac{c d_H \kappa}{3 \pi f} = 1400 \, \mathrm{km} \left( \frac{\kappa}{1 \, \mathrm{Gpc}^{-1}} \right) \left( \frac{100 \, \mathrm{Hz}}{f} \right);
\end{equation}
our 68\% credible bound on $\kappa$ implies $l_0 \lesssim 40 \, \mathrm{km}$ at $100 \, \mathrm{Hz}$.  This compares favorably to the constraint in \citet{Okounkova_2022}, which had $l_0 \lesssim 1000 \, \mathrm{km}$.

\subsubsection{Wang et al.}
\label{sec:comparison_Wang}

\citet{Wang_2021} performed \ac{PE} on GWTC-1 events with a frequency-dependent model of birefringence resembling ours.
Following \cite{Zhao:2019xmm}, that reference parametrized the birefringent waveform in terms of some amplitude and phase modifications to the linear polarizations ($\delta h$ and $\delta \Psi$ respectively), such that
\begin{equation}
    h_{+/\times}^{\rm BR}(f) = h_{+/\times}^{\rm GR}(f)\mp h_{\times/+}^{\rm GR}(f)(i\delta h-\delta\Psi)\,,
\end{equation}
with the plus (minus) sign for the cross (plus) polarization.
In terms of circular polarizations, this is equivalent to 
\begin{align}
h^{\rm BR}_{L/R}(f) &= h^{\rm GR}_{L/R}(f) \left(1 \mp \delta h \mp i \delta \Psi \right) \nonumber\\
&\approx h^{\rm GR}_{L/R}(f) \exp(\mp \delta h \mp i \delta \Psi) \, ,
\end{align}
assuming a small $\delta h$ and $\delta \Psi$ in the last line.
To consider only amplitude birefringence, as in this work, we must compare to the result in Ref.~\citep{Wang_2021} that set $\delta \Psi = 0$ and parametrized $\delta h = \pi f z h_P / M_{\rm PV}$, where $h_P$ is Planck's constant and $M_{\rm PV}$ is the energy scale of the birefringent (parity-violating) correction.%
\footnote{Concretely, \citet{Wang_2021} write $\delta h = - A_\nu \pi f$ with $A_\nu = M_{\rm PV}^{-1} \left[\alpha_\nu (z=0) - \alpha_\nu(z) \left(1+z\right)\right]$, for $\alpha_\nu$ some function of redshift encoding the evolution of a birefringence-mediating field; to produce their constraint, however, they further set $\alpha_\nu(z) = 1$, yielding $A_\nu = -z / M_{\rm PV}$ and hence $\delta h = \pi f z h_P / M_{\rm PV}$, multiplying by $h_P$ to obtain the right dimensions.}
Comparing to our Eq.~\eqref{eq:waveform_modification} and approximating $z \approx d_C H_0/c$ via the Hubble constant, $H_0$, this means that
\begin{equation}
    \label{eq:kappa-to-Mpv}
    |\kappa| \simeq \frac{\pi H_0}{c} \frac{h_P \left( 100 \, \mathrm{Hz} \right)}{M_{\rm PV}} \simeq \frac{3.0 \times 10^{-22} \, \mathrm{GeV}}{M_\mathrm{PV}} \, \mathrm{Gpc}^{-1} .
\end{equation}
\citet{Wang_2021} quote a constraint of $M_{\rm PV} > 10^{-22}\, {\rm GeV}$ at 90\% credibility, which translates into %
  \input{output/kappa_Wang.txt}\unskip\label{output/kappa_Wang.txt}\unskip%
.
We obtained a tighter 90\% upper limit of %
  \input{output/restricted_absolute_kappa_90.txt}\unskip\label{output/restricted_absolute_kappa_90.txt}\unskip%
, or %
  \input{output/M_PV_constraint.txt}\unskip\label{output/M_PV_constraint.txt}\unskip%
.

\subsubsection{Zhu et al.}
\label{sec:comparison_Zhu}

A more recent work by \citet{Zhu:2023wci}, which appeared while this manuscript was finalized, performed \ac{PE} on GWTC-3 events with a frequency-dependent model of amplitude birefringence.
Their model is the same as the one used in \citet{Wang_2021}, which allows us to compare to our results as we did in Sec.~\ref{sec:comparison_Wang}.
\citet{Zhu:2023wci} reported a constraint of $M_{\rm PV} > 4.1 \times 10^{-22}\, {\rm GeV}$ at 90\% credibility, which translates into %
  \input{output/kappa_Zhu.txt}\unskip\label{output/kappa_Zhu.txt}\unskip%
 per Eq.~\eqref{eq:kappa-to-Mpv} above.

\begin{figure}
    \script{posterior_MPV.py}
    \includegraphics[width=\columnwidth]{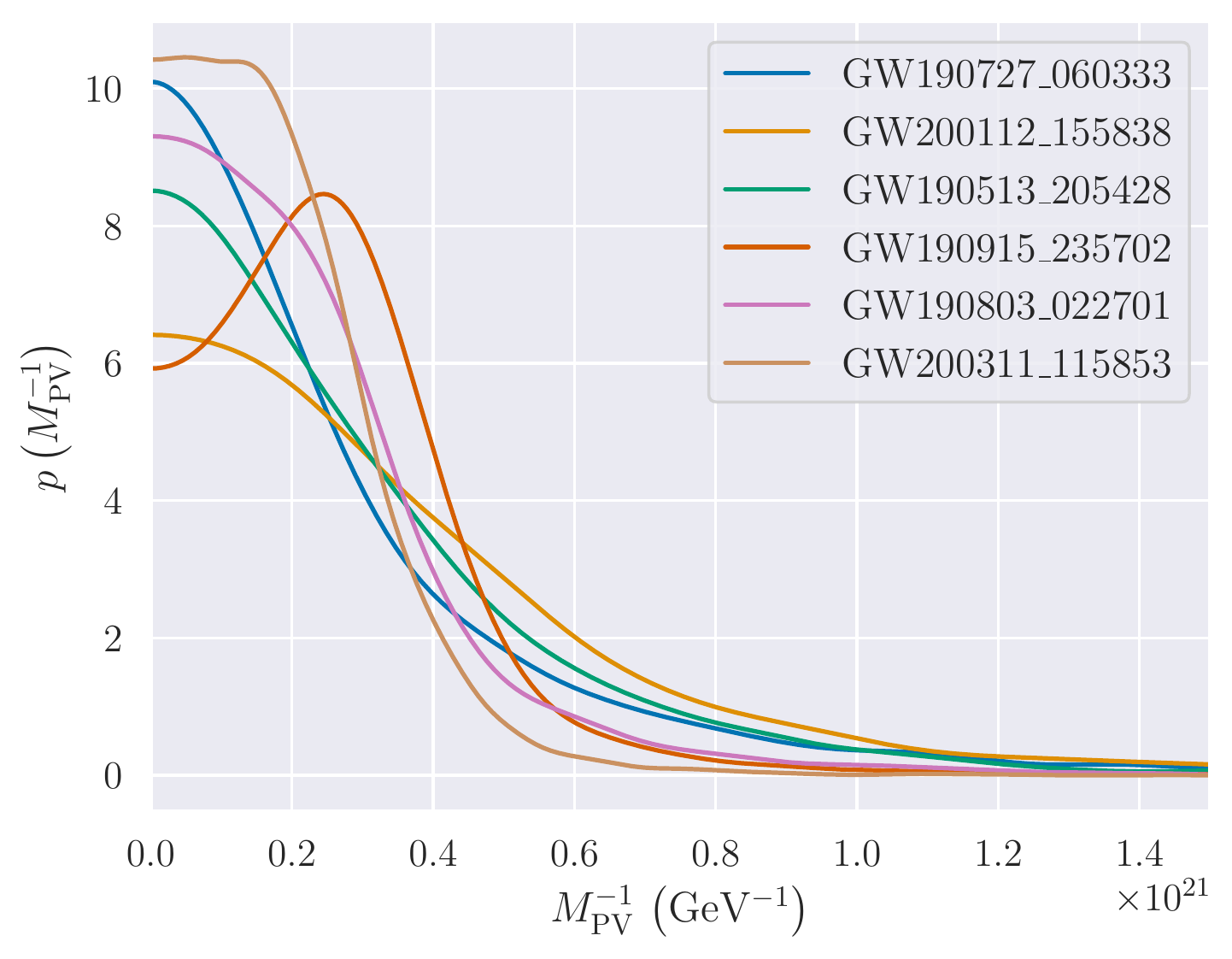}
    \caption{
        Individual-event $M_\mathrm{PV}^{-1}$ posteriors from our analysis, converted from the $\kappa$ posteriors in Fig.~\ref{fig:violin_kappa} via Eq.~\eqref{eq:kappa-to-Mpv}.
        Chosen events are those highlighted in Fig.~1 of \citet{Zhu:2023wci}.
    }
    \label{fig:posterior_MPV}
\end{figure}

This constraint is weaker than ours in spite of the similar number of events considered, a discrepancy that we trace back to inference at the individual-event level.
Figure \ref{fig:posterior_MPV} shows the posteriors on $M_\mathrm{PV}^{-1}$ derived from our analysis via Eq.~\eqref{eq:kappa-to-Mpv} for each event highlighted in Fig.~1 of \citet{Zhu:2023wci}.
Our posteriors differ from those in \citet{Zhu:2023wci} in both overall scale and specific shape.
For example, the posterior for GW190727\_060333 peaks at zero in our analysis, but peaks at a nonzero value in \citet{Zhu:2023wci}.
All distributions are narrower in our analysis by a factor of ${\sim}10\times$.
This might be explained by a number of analysis differences.
The fact that posteriors disagree at the individual-event level suggests that the discrepancy originates in the different choice of parametrization and corresponding priors, in addition to potentially unstated differences in implementation.


\subsection{Parameter degeneracies}
\label{sec:degeneracies}

In examining our results in Sec.~\ref{sec:results:notable}, we identified interactions between birefringence and the inference of source properties like the component spin parameters---in addition to expected partial degeneracies with inclination and distance (Sec.~\ref{sec:inclination}).
In particular, we found that the frequency-dependent modification to the waveform induced by birefringence can effectively mimic spin effects, including precession, especially for heavier systems which are observed over short durations.

\begin{figure}
    \script{mass_ratio.py}
    \includegraphics[width=\columnwidth]{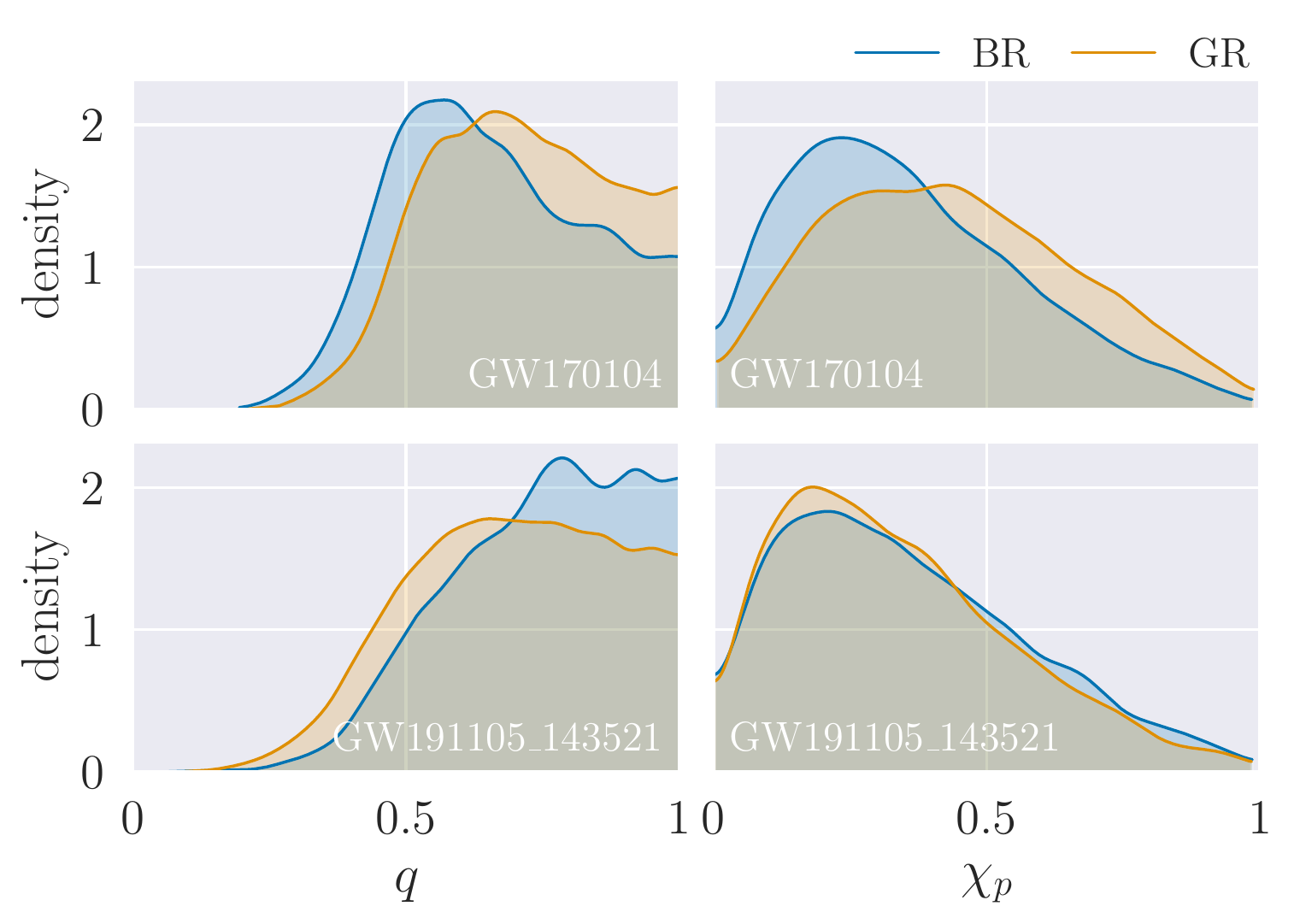}
    \caption{
    Posterior on mass ratio $q \equiv m_2/m_1$ (left) and precessing spin $\chi_p$ (right) for GW170104 (top) and GW191105 (bottom), as obtained from both the birefringent analysis (orange) and the regular \ac{GR} analysis (blue).
    }
    \label{fig:mass_ratio}
\end{figure}

As a consequence, a system that would be inferred to be highly spinning in \ac{GR} may instead be inferred to be low-spinning but highly birefringent if nonzero values of $\kappa$ are allowed.
These morphological degeneracies can interact with the distribution of prior probability mass to yield a preference for nonzero $\kappa$ in some events: in absence of a strong constraint from the data, it can be more favorable to place a source at far distances with large $|\kappa|$, than at close distances with a fine tuned spin configuration and low $|\kappa|$.
The resulting posterior can be unimodal or bimodal depending on the specific phase and polarization states observed at the detector.
The events GW170818 (Sec.~\ref{sec:GW170818}) and GW190521 (Sec.~\ref{sec:GW190521}) appear to fall into these two categories, respectively.

Conversely, we expect that support for birefringence can also appear for signals that do not necessarily show significant spin effects in the \ac{GR} analysis, as long as the spins or other source parameters can be tuned to \emph{counteract} the frequency-dependent dephasing that would be otherwise induced by a nonzero $\kappa$.
In those situations, $\kappa$ may become at least partially degenerate with inclination, also leading to a bimodal posterior.
This is because the intrinsic source parameters may, to an extent, ``absorb'' the frequency dependence of the birefringent effect, approximately reinstating the degeneracy expected for a single Fourier mode (Sec.~\ref{sec:inclination}).

Besides spins, other parameters can interact with $\kappa$ either directly or through their own couplings with the spins.
Since the mass ratio can affect the phasing of a \ac{GW} inspiral at similar order as the spins \cite{Blanchet:2013haa}, we might expect some degree of interaction between $q$ and $\kappa$.
This seems to indeed be the case for some events, like GW170104 or GW191105, as we demonstrate in Fig.~\ref{fig:mass_ratio}.
In the former, introducing birefringence increases the preference for asymmetric masses ($q < 1$) while decreasing the inferred level of precession; in the latter, birefringence increases the preference for symmetric masses ($q = 1$) but does not affect the inference on precession.
The $\kappa$ posteriors are bimodal for both events, but do not rule out $\kappa = 0$ (Fig.~\ref{fig:violin_kappa}).
In neither case is $\chi_{\rm eff}$ significantly impacted by birefringence.
The coupling between $q$ and $\kappa$ is likely limited by the fact that changes in the mass ratio alone cannot produce significant amplitude modulations.

Since our inference on $\kappa$ can be tied to precession and the spin orientations, we might be susceptible to systematics in the modeling of spin angles in \textsc{IMRPhenomXPHM} and might thus benefit from further analysis with other waveforms like \textsc{NRSur7dq4} \cite{Varma:2018mmi}.

\section{Conclusion}
\label{sec:Conclusion}

We have reanalyzed all \acp{BBH} in GWTC-3 with $\mathrm{FAR} < 1/\mathrm{yr}$ to constrain amplitude birefringence in the propagation of \acp{GW}.
To this end, we implemented a model in which right or left handed polarizations are amplified or suppressed over distance as a function of frequency, with an overall strength parametrized by a birefringent ``attenuation'' parameter $\kappa$, following Eq.~\eqref{eq:waveform_modification}.
This parametrization is consistent with parity-odd theories like Chern-Simons gravity and can be used to constrain them where applicable.

We found no evidence of amplitude birefringence in the GWTC-3 data and constrained %
  \input{output/restricted_kappa_median.txt}\unskip\label{output/restricted_kappa_median.txt}\unskip%
with 90\% credibility when treating $\kappa$ as a global quantity shared by all GWTC-3 events (Sec.~\ref{sec:results:gwtc}).
This measurement is significantly more stringent than past constraints (Table \ref{tab:comparison_summary}).
As an additional check, we implemented a hierarchical analysis that allowed for each event to probe different effective values of $\kappa$, as might be the case if birefringence is mediated by a field that is nonuniform over the angular separations between detected sources (Sec.~\ref{sec:results:hier}).
The result of that analysis was consistent with a vanishing $\kappa$ for all events in our catalog within 90\% credibility, but hinted at some possible variance in the population.
Future catalogs with more events will shed light on this feature and enable richer models with explicit correlations between birefringent attenuation and source parameters like sky location \citep{Goyal:2023uvm,Ezquiaga:2021ler}.

From the set of results from individual events, we highlighted two, GW170818 and GW190521, whose posterior manifests interactions between birefringence and our inference of source parameters (Sec.~\ref{sec:results:notable}).
In particular, we identified the relevance of spins and their (partial) degeneracy with $\kappa$, in addition to expected correlations with source inclination and distance.
The mass ratio can also play a role, by coupling with the spins or $\kappa$ directly (Sec.~\ref{sec:degeneracies}).

Motivated by Chern-Simons gravity, this study only focused on amplitude birefringence.
Other parity-violating gravity theories also predict velocity birefringence, which can dominate over amplitude birefringence when both are present \cite{Zhao:2019xmm}.
Velocity birefringence was tested with GWTC-3 in \citet{Wang:2021gqm} and \citet{Haegel:2022ymk}.

This study was restricted to \acp{BBH} because of the expectation that they should dominate the birefringence constraint thanks to their larger redshifts.
However, lower mass systems involving neutron stars may also be informative thanks to the wide band of frequencies spanned by their signals, in spite of their closer distances.
Analysis of future catalogs will benefit from the inclusion of those events, as well as a much larger number of \ac{BBH} sources at increasingly greater distances.
Future measurements will also be enriched by the addition of KAGRA \cite{KAGRA} and LIGO India to the \ac{GW} detector network, which will allow us to better disambiguate between polarization states and hence between models of birefringent propagation.
All our data are available in \citet{dataset}.

\begin{acknowledgments}
We thank Macarena Lagos and Nicol\'as Yunes for helpful discussions.
M.I., K.W.K.W.~, and W.M.F.~ are funded by the Center for Computational Astrophysics at the Flatiron Institute.
The Flatiron Institute provided the computational resources used in this work.
This research has made use of data or software obtained from the Gravitational Wave Open Science Center, a service of LIGO Laboratory, the LIGO Scientific Collaboration, the Virgo Collaboration, and KAGRA.
LIGO Laboratory and Advanced LIGO are funded by the United States National Science Foundation (NSF) as well as the Science and Technology Facilities Council (STFC) of the United Kingdom, the Max-Planck-Society (MPS), and the State of Niedersachsen/Germany for support of the construction of Advanced LIGO and construction and operation of the GEO600 detector.
Additional support for Advanced LIGO was provided by the Australian Research Council.
Virgo is funded, through the European Gravitational Observatory (EGO), by the French Centre National de Recherche Scientifique (CNRS), the Italian Istituto Nazionale di Fisica Nucleare (INFN) and the Dutch Nikhef, with contributions by institutions from Belgium, Germany, Greece, Hungary, Ireland, Japan, Monaco, Poland, Portugal, Spain.
KAGRA is supported by Ministry of Education, Culture, Sports, Science and Technology (MEXT), Japan Society for the Promotion of Science (JSPS) in Japan; National Research Foundation (NRF) and Ministry of Science and ICT (MSIT) in Korea; Academia Sinica (AS) and National Science and Technology Council (NSTC) in Taiwan.
This paper was compiled using \textsc{showyourwork} \cite{Luger2021} to facilitate reproducibility.
\end{acknowledgments}

\appendix*

\section{Extended results for notable events}
\label{sec:Appendix}

\subsection{GW170818}
\label{sec:corner_GW170818_appendix}

\begin{figure*}[h]
    \script{corner_GW170818_appendix.py}
    \includegraphics[width=\textwidth]{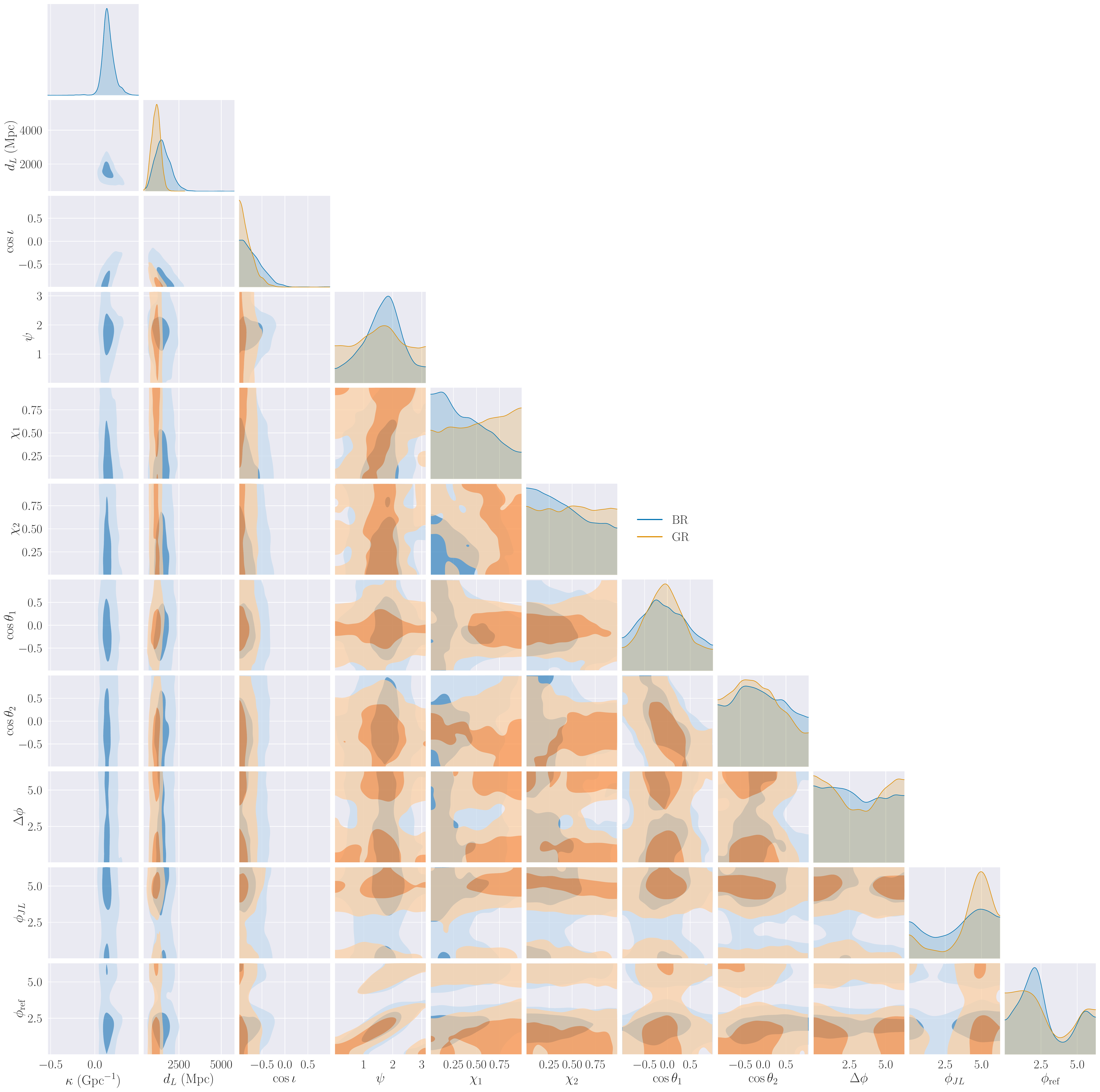}
    \caption{
        Extended corner plot for GW170818: a supplement to Fig.~\ref{fig:corner_GW170818} discussed in Appendix~\ref{sec:corner_GW170818_appendix}.
        The shaded regions contain 90\% and 39.35\% ($1\sigma$) of the probability mass.
        The prior is uniform in all shown quantities except $d_L$, whose prior corresponds to a distribution uniform in comoving volume.
    }
    \label{fig:corner_GW170818_appendix}
\end{figure*}

Here we provide further details on the GW170818 measurement. 
Figure \ref{fig:corner_GW170818_appendix} displays the posterior on all parameters we consider relevant for this event, of which Fig.~\ref{fig:corner_GW170818} in the main text represents a subset.
As in that figure, Fig.~\ref{fig:corner_GW170818_appendix} shows the regular \textsc{IMRPhenomXPHM} \ac{GR} analysis (GR; orange) and the birefringent analysis (BR; blue), both of which show notable features.

The GR analysis stands out for its relatively confident identification of the spin angles, $\theta_{1/2}$, $\Delta \phi$, and $\phi_{JL}$, as well as the phase and polarization angles, $\phi_{\rm ref}$ and $\psi$.
Of the former, the two $\theta_{1/2}$ angles encode the tilts of the component \acp{BH} with respect to the orbital angular momentum $\vec{L}$, and $\Delta \phi$ is the angle between the spin projections onto the orbital plane, while $\phi_{JL}$ is a similar angle separating the projections of $\vec{L}$ and the total angular momentum $\vec{J}$;
of the latter, $\phi_{\rm ref}$ is an overall reference phase, and $\psi$ encodes the orientation of the binary within the plane of the sky \cite{Isi:2022mbx}.
All these parameters are anchored to a reference point in the inspiral, which in this case corresponds to the time at which the dominant multipole of the observed \ac{GW} signal reaches 20 Hz at the detector (spin angles may be better identified by using a more physical reference point \cite{Varma:2021csh}).
It is unusual for all these angles to be well constrained, which suggests that data for this event display a particular phase and polarization signature.

The joint posterior on the spin magnitudes, $\chi_1$ and $\chi_2$, disfavors the origin $\chi_1 = \chi_2 = 0$ with high credibility, indicating that at least one of the component \acp{BH} must have been highly spinning (this is not apparent from the marginals because the \acp{BH} were equal in mass \cite{Biscoveanu:2020are}).
Furthermore, the fact that the tilts are favored to be close to $\theta_{1/2} \approx \pi/2$ implies that the spins must lie along the orbital plane;
this is a restatement of the support for high $\chi_p$ in Fig.~\ref{fig:corner_GW170818}.

On the other hand, the birefringent analysis does not strongly favor high spins; this is evidenced by the joint posterior on $\chi_1$ and $\chi_2$, which now favors $\chi_1 = \chi_2 = 0$ and assigns relatively low probability densities to higher values.
Although the spin tilts are similarly constrained to $\theta_{1/2} \approx \pi/2$, the low spin magnitudes would imply weak if any precession, which is reflected in the support for lower values of $\chi_p$ in Fig.~\ref{fig:corner_GW170818} of the main text.
As discussed in Sec.~\ref{sec:GW170818}, this can be explained by the introduction of $\kappa$, which can result in morphologies that are approximately degenerate with precession over a small number of waveform cycles (cf.~Fig.~\ref{fig:birefringence}).
See the main text for further discussion.

\subsection{GW190521}
\label{sec:corner_GW190521_appendix}

\begin{figure*}[h]
    \script{corner_GW190521_appendix.py}
    \includegraphics[width=\textwidth]{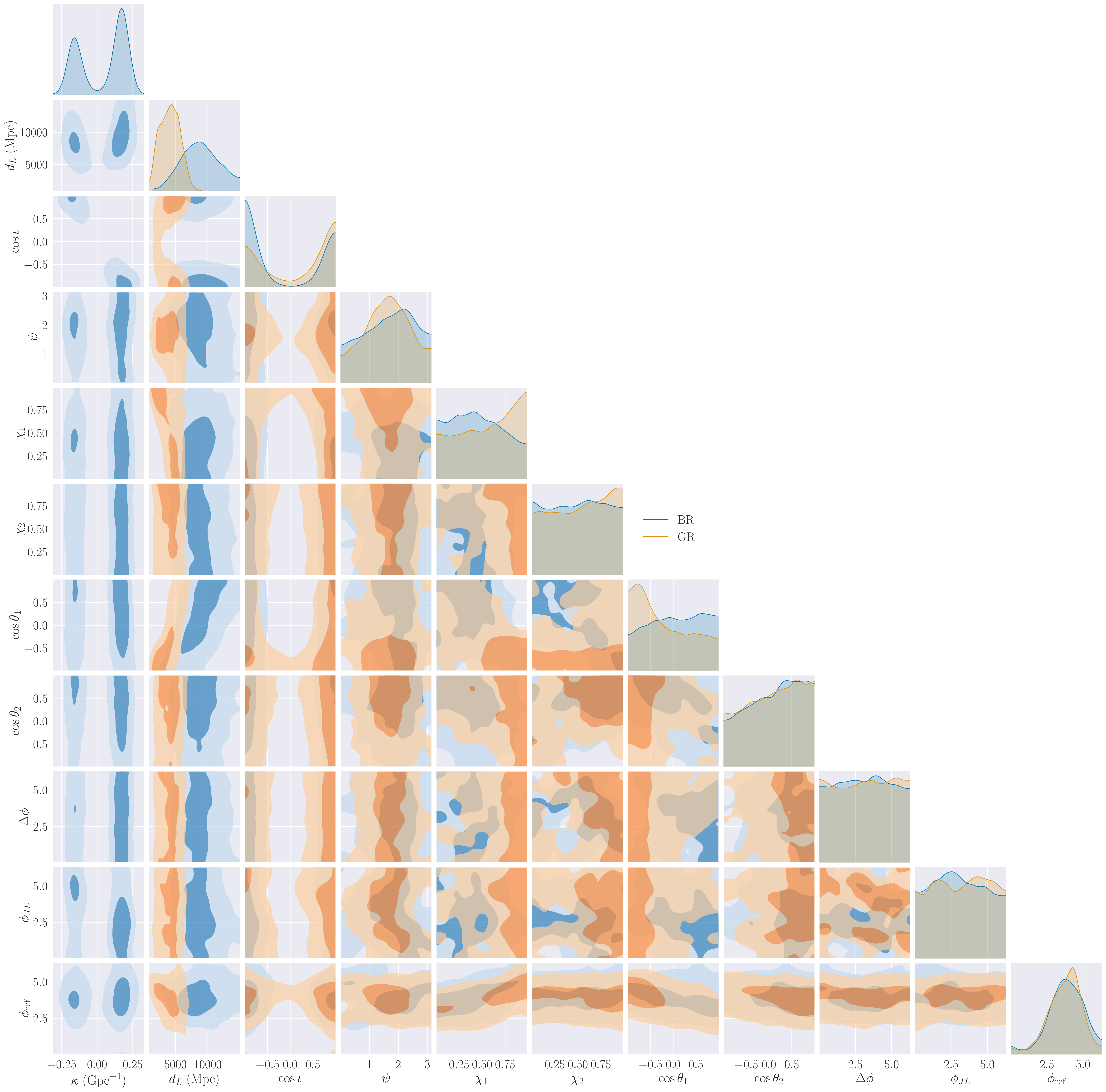}
    \caption{
        Extended corner plot for GW190521: a supplement to Fig.~\ref{fig:corner_GW190521} discussed in Appendix~\ref{sec:corner_GW190521_appendix}.
        The shaded regions contain 90\% and 39.35\% ($1\sigma$) of the probability mass.
        The prior is uniform in all shown quantities except $d_L$, whose prior corresponds to a distribution uniform in comoving volume.
    }
    \label{fig:corner_GW190521_appendix}
\end{figure*}

As in the previous section, here we display additional parameters from the GW190521 measurement in Fig.~\ref{fig:corner_GW190521_appendix}, expanding upon Fig.~\ref{fig:corner_GW190521} in the main text.
The reference \ac{GR} analysis, obtained with the \textsc{IMRPhenomXPHM} waveform, favors near extremal spins, in particular, for the heavier component, and rules out nonspinning objects (i.e., $\chi_1 = \chi_2 = 0$) with $\geq 90\%$ credibility.
This can be seen most clearly in the joint posterior for $\chi_1$ and $\chi_2$, rather than in the respective 1D marginals because the \acp{BH} in this system were inferred to have equal masses \cite{Biscoveanu:2020are}.

Similar to GW170818, the \ac{GR} analysis for GW190521 also provides an informative measurement of the polarization angle $\psi$ and reference phase $\phi_{\rm ref}$.
However, the fact that $\psi$ is constrained does not alone imply that the degeneracy between right- and left-handed states is broken (for a nonprecessing source, $\psi$ is related to the difference in phase between the circular polarization modes \cite{Isi:2022mbx}).
Additionally, the posterior for tilt angle of the primary \ac{BH} $\theta_1$ shows support for an antialigned spin ($\cos\theta_1 \approx -1$).
For these events, spin and phase angles are referred to 10 Hz at the detector.

Unlike the \ac{GR} analysis, the birefringent analysis does not favor extremal spins and is, in fact, consistent with nonspinning objects within 90\% credibility (lighter blue region encloses $\chi_1 = \chi_2 = 0$ in the respective panel of Fig.~\ref{fig:corner_GW190521_appendix}).
In Fig.~\ref{fig:corner_GW190521}, this same fact manifested as a $\chi_{\rm eff}$ posterior with higher support for the origin.
As we discussed in Sec.~\ref{sec:GW190521}, this suggests that some of the spin effects in these data are being absorbed by a nonzero $\kappa$.
Other than the spins, the birefringent analysis differs most notably from the \ac{GR} one in the inference of the luminosity distance, which is favored to be much greater.
See Sec.~\ref{sec:GW190521} in the main text for related exposition.

\bibliography{bib}

\end{document}

%% file: output/restricted_kappa_median.txt
$\kappa = -0.019^{+0.038}_{-0.029} \, \mathrm{Gpc}^{-1}$

%% file: output/M_PV_constraint.txt
$M_{\rm PV} \gtrsim 6.8 \times 10^{-21}\, {\rm GeV}$

%% file: output/CL_kappa_0.txt
$0.209$

%% file: output/GW170818_constraint.txt
$\mu_i / \sigma_i = 2.0$

%% file: output/GW200129_constraint.txt
$\mu_i / \sigma_i = -1.5$

%% file: output/best_events_kappa.txt
\begin{tabular}{lccc}Event & $\kappa$ ($\mathrm{Gpc}^{-1}$) & $\sigma_i$ ($\mathrm{Gpc}^{-1}$) & CL\\ \hline GW200129\_065458 & $-0.072^{+0.085}_{-0.071}$ & $0.048$ & $0.876$\\GW200224\_222234 & $-0.069^{+0.086}_{-0.071}$ & $0.058$ & $0.860$\\GW200311\_115853 & $-0.036^{+0.115}_{-0.077}$ & $0.061$ & $0.584$\\GW190412 & $-0.091^{+0.129}_{-0.086}$ & $0.068$ & $0.857$\\GW191204\_171526 & $+0.004^{+0.100}_{-0.109}$ & $0.069$ & $0.205$\\GW190512\_180714 & $+0.002^{+0.113}_{-0.109}$ & $0.070$ & $0.005$\\GW170818 & $+0.146^{+0.136}_{-0.088}$ & $0.076$ & $0.977$\\GW200219\_094415 & $-0.001^{+0.125}_{-0.118}$ & $0.079$ & $0.109$\\GW190701\_203306 & $-0.041^{+0.158}_{-0.086}$ & $0.084$ & $0.587$\\GW190413\_052954 & $+0.008^{+0.137}_{-0.113}$ & $0.087$ & $0.081$ \end{tabular}

%% file: output/bimodal_events_mass.txt
\begin{tabular}{lcccc}Event & $M$ ($M_{\odot}$) & $\chi_p$ & $\chi_{\rm eff}$ & CL\\ \hline GW170104 & $60.6^{+4.0}_{-3.9}$ & $0.42^{+0.41}_{-0.32}$ & $-0.04^{+0.15}_{-0.20}$ & $0.786$\\GW190413\_134308 & $133.3^{+20.6}_{-20.5}$ & $0.56^{+0.35}_{-0.41}$ & $-0.04^{+0.28}_{-0.36}$ & $0.837$\\GW190521 & $250.1^{+40.3}_{-37.5}$ & $0.46^{+0.36}_{-0.32}$ & $+0.04^{+0.29}_{-0.39}$ & $0.996$\\GW190805\_211137 & $146.6^{+19.8}_{-20.1}$ & $0.53^{+0.33}_{-0.33}$ & $+0.33^{+0.29}_{-0.36}$ & $0.895$\\GW191105\_143521 & $22.4^{+2.6}_{-0.6}$ & $0.31^{+0.44}_{-0.24}$ & $-0.03^{+0.11}_{-0.08}$ & $0.802$ \end{tabular}

%% file: output/mu_median.txt
$\mu = -0.008^{+0.031}_{-0.031} \, \mathrm{Gpc}^{-1}$

%% file: output/sigma_median.txt
$\sigma < 0.048 \, \mathrm{Gpc}^{-1}$

%% file: output/generic_kappa_median.txt
$\kappa_i = -0.011^{+0.065}_{-0.051} \, \mathrm{Gpc}^{-1}$

%% file: output/comparison_summary.txt
\begin{tabular}{lccc} & $M_{\rm PV}$ ($10^{-21}\, {\rm GeV}$) & $|\kappa|$ ($\mathrm{Gpc}^{-1}$) & CL (\%) \\ \hline \citet{Wang_2021} & $> 0.10$ & $< 2.94$ & 90 \\ \citet{Okounkova_2022} & $> 0.40$ & $< 0.74$ & 68 \\ \citet{Zhu:2023wci} & $> 0.41$ & $< 0.72$ & 90 \\ Ng \emph{et al.} (this work) & $> 6.75$ & $< 0.04$ & 90 \end{tabular}

%% file: output/restricted_absolute_kappa_68.txt
$|\kappa| < 0.03 \, \mathrm{Gpc}^{-1}$

%% file: output/improvement_Okounkova.txt
${\sim}25\times$

%% file: output/kappa_Wang.txt
$|\kappa| \lesssim 2.94 \, \mathrm{Gpc}^{-1}$

%% file: output/restricted_absolute_kappa_90.txt
$|\kappa| < 0.04 \, \mathrm{Gpc}^{-1}$

%% file: output/kappa_Zhu.txt
$|\kappa| \lesssim 0.72 \, \mathrm{Gpc}^{-1}$